\documentclass[sigconf]{acmart}
\usepackage{fancybox}
\usepackage{multirow}
\usepackage{xspace}
\usepackage{framed}
\usepackage{threeparttable}
\usepackage{subfigure}
\usepackage{listings}

\AtBeginDocument{%
  \providecommand\BibTeX{{%
    \normalfont B\kern-0.5em{\scshape i\kern-0.25em b}\kern-0.8em\TeX}}}

\copyrightyear{2022}
\acmYear{2022}
\setcopyright{acmcopyright}\acmConference[ICPC '22]{30th International Conference
on Program Comprehension}{May 16--17, 2022}{Virtual Event, USA}
\acmBooktitle{30th International Conference on Program Comprehension (ICPC '22),
May 16--17, 2022, Virtual Event, USA}
\acmPrice{15.00}
\acmDOI{10.1145/3524610.3527888}
\acmISBN{978-1-4503-9298-3/22/05}
\begin{document}

\newcommand{\ie}{\textit{i}.\textit{e}.}
\newcommand{\eg}{\textit{e}.\textit{g}.}

\newcommand{\todoc}[2]{{\textcolor{#1}{\textbf{[#2]}}}}
\newcommand{\todoblue}[1]{\todoc{blue}{\textbf{#1}}}
\newcommand{\todored}[1]{\todoc{red}{\textbf{#1}}}
\newcommand{\todoorange}[1]{\todoc{orange}{\textbf{#1}}}
\newcommand{\gu}[1]{\todoblue{Gu: #1}}
\newcommand{\shen}[1]{\todoorange{Shen: #1}}
\newcommand{\cn}[1]{\todored{Cui: #1}}
\newcommand{\jyz}[1]{\todored{Jyz: #1}}
\newcommand{\approach}{Zecoler\xspace}

%%
%% The "title" command has an optional parameter,
%% allowing the author to define a "short title" to be used in page headers.
\title{Zero-Shot Program Representation Learning
}

%%
%% The "author" command and its associated commands are used to define
%% the authors and their affiliations.
%% Of note is the shared affiliation of the first two authors, and the
%% "authornote" and "authornotemark" commands
%% used to denote shared contribution to the research.

\author{Nan Cui, Yuze Jiang, Xiaodong Gu, Beijun Shen}
\authornote{Beijun Shen is the corresponding author.}
\affiliation{%
  \institution{School of Software, Shanghai Jiao Tong University}
  \city{Shanghai}
  \country{China}
}
\email{{cuinan, jyz-1201, xiaodong.gu, bjshen}@sjtu.edu.cn}

\begin{comment}
\author{Nan Cui}
\authornote{Both authors contributed equally to this research.}
\email{trovato@corporation.com}
\orcid{1234-5678-9012}
\author{G.K.M. Tobin}
\authornotemark[1]
\email{webmaster@marysville-ohio.com}
\affiliation{%
  \institution{Institute for Clarity in Documentation}
  \streetaddress{P.O. Box 1212}
  \city{Dublin}
  \state{Ohio}
  \country{USA}
  \postcode{43017-6221}
}

\author{Lars Th{\o}rv{\"a}ld}
\affiliation{%
  \institution{The Th{\o}rv{\"a}ld Group}
  \streetaddress{1 Th{\o}rv{\"a}ld Circle}
  \city{Hekla}
  \country{Iceland}}
\email{larst@affiliation.org}
\end{comment}

%%
%% By default, the full list of authors will be used in the page
%% headers. Often, this list is too long, and will overlap
%% other information printed in the page headers. This command allows
%% the author to define a more concise list
%% of authors' names for this purpose.
\renewcommand{\shortauthors}{Nan Cui, et al.}

%%
%% The abstract is a short summary of the work to be presented in the
%% article.
\begin{abstract}
Learning program representations has been the core prerequisite of code intelligence tasks (\eg, code search and code clone detection). The state-of-the-art pre-trained models such as CodeBERT require the availability of large-scale code corpora. However, gathering training samples can be costly and infeasible for domain-specific languages such as Solidity for smart contracts. In this paper, we propose \approach, a zero-shot learning approach for code representations. \approach is built upon a pre-trained programming language model. In order to elicit knowledge from the pre-trained models efficiently, \approach casts the downstream tasks to the same form of pre-training tasks by inserting trainable prompts into the original input. 
Then, it employs the prompt learning technique to optimize the pre-trained model by merely adjusting the original input. This enables the representation model to efficiently fit the scarce task-specific data while reusing pre-trained knowledge.
%In order to efficiently fit the scarce task-orient data while reusing pre-trained knowledge, \approach employs prompt-based learning paradigm which optimizes the PLM by merely adjusting the original input.
We evaluate \approach in three code intelligence tasks in two programming languages that have no training samples, namely, Solidity and Go, with model trained in corpora of common languages such as Java. Experimental results show that our approach significantly outperforms baseline models in both zero-shot and few-shot settings. 
\end{abstract}

%%
%% The code below is generated by the tool at http://dl.acm.org/ccs.cfm.
%% Please copy and paste the code instead of the example below.
%%
% \begin{CCSXML}
% <ccs2012>
%  <concept>
%   <concept_id>10010520.10010553.10010562</concept_id>
%   <concept_desc>Computer systems organization~Embedded systems</concept_desc>
%   <concept_significance>500</concept_significance>
%  </concept>
%  <concept>
%   <concept_id>10010520.10010575.10010755</concept_id>
%   <concept_desc>Computer systems organization~Redundancy</concept_desc>
%   <concept_significance>300</concept_significance>
%  </concept>
%  <concept>
%   <concept_id>10010520.10010553.10010554</concept_id>
%   <concept_desc>Computer systems organization~Robotics</concept_desc>
%   <concept_significance>100</concept_significance>
%  </concept>
%  <concept>
%   <concept_id>10003033.10003083.10003095</concept_id>
%   <concept_desc>Networks~Network reliability</concept_desc>
%   <concept_significance>100</concept_significance>
%  </concept>
% </ccs2012>
% \end{CCSXML}

% \ccsdesc[500]{Computer systems organization~Embedded systems}
% \ccsdesc[300]{Computer systems organization~Redundancy}
% \ccsdesc{Computer systems organization~Robotics}
% \ccsdesc[100]{Networks~Network reliability}

%%
%% Keywords. The author(s) should pick words that accurately describe
%% the work being presented. Separate the keywords with commas.

%% A "teaser" image appears between the author and affiliation
%% information and the body of the document, and typically spans the
%% page.
% \begin{teaserfigure}
%   \includegraphics[width=\textwidth]{sampleteaser}
%   \caption{Seattle Mariners at Spring Training, 2010.}
%   \Description{Enjoying the baseball game from the third-base
%   seats. Ichiro Suzuki preparing to bat.}
%   \label{fig:teaser}
% \end{teaserfigure}

%%
%% This command processes the author and affiliation and title
%% information and builds the first part of the formatted document.
\maketitle

\section{Introduction}
\label{sec:intro}
% \cn{What is the difference between representation learning and adapting PLMs to downstream tasks? explain the difference.}
% \cn{more innovation: continuous vectors extract code and code task knowledge}
Learning program representations has achieved great success in software engineering thanks to recent advances in deep learning~\cite{BengioCV13} and the availability of large-scale code corpora~\cite{gpt3}.  
Program representations, namely, code vectors that reflect their deep semantics, have been widely used in code intelligence tasks \cite{codexglue} such as clone detection~\cite{FangLS0S20}, code summarization~\cite{ChoiBNL21}, and code search~\cite{HaldarWXH20}. For example, in the clone detection task, program representations can be used to reflect similar features between two code snippets~\cite{ZhangHZWLS21}.
%, through non-AI tools based on some artificially created rules. Although these tools can barely complete tasks, ``AI for Code" can be re-tuned and adapt for more tasks automatically, resolve code in various format robustly and dig into features in code that human hard to notice \cite{codenet}. Deep Learning (DL) is used to implement AI for code frequently, and many DL models are customized to support code tasks \cite{MastropaoloSCNP21}.
%Learning program representation is the first procedure of AI for code models. A good program representation contains semantic information of code for models to understand code snippet. Traditional models cost big dataset related with code and downstream tasks to train program representation \cite{ZhangHZWLS21}.

Despite showing promising results, the state-of-the-art techniques rely on the availability of large code corpora. However, labeling data samples of some programming languages is often expensive and sometimes infeasible. For example, Solidity is a new language and specifically designed for smart contracts. This language is becoming increasingly popular and hence code representation tools for the Solidity language are highly demanded. However, obtaining labelled Solidity code is challenging as it requires domain knowledge on Blockchain and the collected data is significantly redundant~\cite{ChenLZ0Z21}. This restricts the collection of supervised data and causes deep learning models to learn poor representations~\cite{codenet}. %\cn{There are many other languages may lack of labeled data, even general languages like Java and Go can have insufficient training data for some downstream tasks.}

One possible solution towards alleviating this issue is to use the pre-trained language models (PLMs) such as CodeBERT~\cite{codebert}, PL-BART~\cite{plbart}, and CodeT5~\cite{codet5}. PLMs are pre-trained to learn code representations of large common languages such as Java and then are fine-tuned on domain-specific languages such as Solidity. For example, \citet{salza2021effectiveness} proposed cross-language CodeBERT, which pre-trains CodeBERT on multiple languages and then transfers the program representations to other languages through fine-tuning. %There are several methods existing that can implement zero-shot learning, such as plain language model (LM) with fine-tuning \cite{codebert} and meta-learning \cite{ChenG0QQWL21}.

However, fine-tuning a PLM on a specific task is challenging. The learning tasks (e.g., masked language model (MLM)) in the pre-training phase are usually different from the downstream tasks (e.g, code search). %data used for fine-tuning is usually different from the data used for pre-training. 
As such, the reusability of prior knowledge learned in the pre-training phase may be limited in the fine-tuning phase. This is even more challenging when there is no or insufficient training data for downstream tasks in domain-specific languages. The large pre-trained model can easily overfit scarce data, which leads to poor task fitting.

%\smallskip\noindent\textbf{Our work.} 
In this paper, we propose \approach (\textbf{Ze}ro-shot \textbf{co}de representation \textbf{le}a\textbf{r}ning), a novel approach for learning program representations without labelled data samples of the target language. In order to learn better representations and bridge the gap between pre-training and downstream tasks, we adapt prompt learning~\cite{abs210713586}, a new learning paradigm for PLMs. Prompt learning adapts the downstream task to the same form as that in pre-training through accompanying trainable prompt tokens with the PLM input. %The trainable prompt embeddings are optimized while keeping the PLM parameters frozen. \cn{We do not freeze PLM in current paper's experiments.} 
The continuous vectors of prompts guide the pre-trained model to elicit knowledge of programming languages efficiently.
For example, the input of code clone detection can be converted to an MLM task input based on a template containing prompt tokens. In this way the model can be adapted to the target language while maximizing the use of prior knowledge learned during the pre-training phase. Hence, prompt learning allows few-shot or even zero-shot learning for pre-trained models in new languages with unlabelled data~\cite{pet}. %and it has been demonstrated to be substantially effective over fine-tuning in many tasks~\cite{}. 

To evaluate the proposed approach, we experiment on three code intelligence tasks, including code clone detection, code search, and method name prediction. The results show that our approach is substantially effective in zero-shot learning of program representations. The accuracy of the three tasks in Solidity is 79.8\%, 67.1\%, and 68.1\%, respectively, which is around 14.7\% greater than the strong baseline CodeBERT. \approach also demonstrates great effectiveness when a few data samples are given. Moreover, the learned program representations have good generalizability.

The contributions of this paper can be summarized as:
\begin{itemize}
    \item To the best of our knowledge, we are the first to propose zero-shot learning of program representations, which does not require manual labeling of data for code intelligence tasks.
    \item We propose a prompt learning based method for zero-shot program representation. 
    \item We conduct extensive experiments to evaluate the proposed approach. Results show that our approach significantly outperforms the baseline models. 
\end{itemize}

\section{Background}
\label{sec:background}
\subsection{Pre-trained Language Models for Code}
\label{sec:pretrain}

Pre-trained language models (PLMs) such as BERT~\cite{DevlinCLT19}, GPT~\cite{radford2018improving}, and T5~\cite{RaffelSRLNMZLL20} have achieved a great success in the natural language processing field. %PLMs refer to deep neural networks that learn common representations of languages from large-scale text corpora.
A PLM is usually pre-trained on a large-scale text corpora through a series of self-supervised learning tasks such as masked language modeling (MLM) and next sentence prediction (NSP). The MLM task masks a random portion of tokens in the input text and tries to predict the masked words, while the NSP task predicts whether or not two given input segments are coherent. 
The PLM can then be fine-tuned on task-specific datasets. A fine-tuning header on top of the PLM is optimized via supervised learning tasks in a specific domain. 

Due to the great success of PLMs, researchers also seek the adaptations of PLMs for programming languages~\cite{codebert,codet5}.
They customize the pre-training objectives using programming related tasks on a big code corpus. PLMs have been successfully used to learn program representations and further be used in a variety of code intelligence tasks.

\begin{figure}
    \centering
    \includegraphics[width=0.5\textwidth, trim=0 70 0 70,clip]{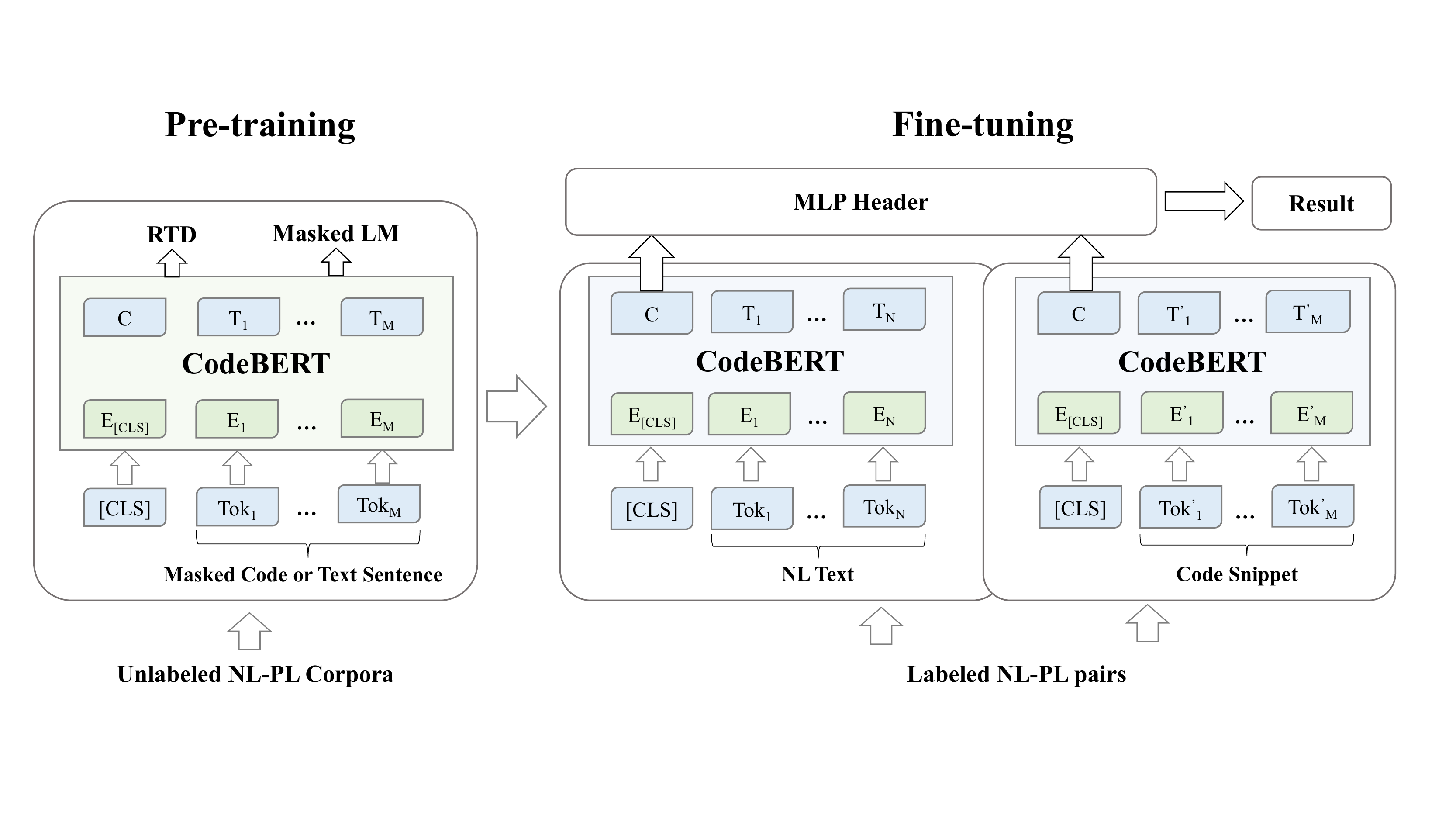}
    \caption{Illustration of the pre-trained code model.}
    \label{fig:plm}
\vspace{-4mm}    
\end{figure}

For example, CodeBERT is built based on RoBERTa~\cite{roberta} and is pre-trained with both natural and programming languages. Figure~\ref{fig:plm} illustrates the main pipeline of CodeBERT. The model is first pre-trained on two tasks, namely, MLM and replaced token detection (RTD). In MLM, CodeBERT randomly masks the tokens in natural language and programming language (NL-PL) pairs and learns to predict the original words. For RTD, the model is trained to detect whether tokens are original or not. 
The pre-trained model is then fine-tuned on data of downstream tasks such as clone detection and code search. A header based on multilayer perceptron (MLP) is added to the PLM and is optimized with downstream tasks. % This header can be a multi-layer perceptron (MLP) neural network or LSTM decoder, as long as it can adapt PLM to the final answer of downstream tasks and do the back propagation to train itself and fine-tune the PLM. 

\subsection{Zero-Shot Learning}
The standard supervised learning approaches train a model with large-scale labelled samples. However, in many tasks such as recognizing name of a new brand or translating a new language, obtaining sufficient training samples is laborious and often impracticable.
Zero-shot learning transfers a learned model to a target domain that has no labelled data, and hence alleviates this ``data hungry'' problem. 
%Zero-shot learning refers to a machine learning paradigm that transfers a learned model to a target domain that has no labelled data. Deep learning is known as ``data hungry'', that means, it often requires large-scale data samples for training the model. However, in many tasks such as recognizing name of a new brand or translating a new language, obtaining sufficient training samples is quite laborious and often impracticable. In this case, zero-shot learning can alleviate the ``data hungry'' problem.
It can be realized through a variety of techniques such as data augmentation~\cite{BorneaPRFS21}, meta-learning~\cite{metalearning1, metalearning2}, and PLMs~\cite{gpt3}.

\textit{Data augmentation.}
A direct technique of zero-shot learning is data augmentation, namely, enlarging the data set (e.g., randomly inserting samples and noise) so that the model can have sufficient data samples for training~\cite{BorneaPRFS21}. 

\textit{Meta-learning.} 
%Meta learning is also known as ``learning to learn'', such as ``N-way-K-shot'' \cite{WangYKN20}, which makes models classify N labels task with only K data samples for training. The scale of data samples, K, can be very small, compared with the large scale of training data in fine-tuning.
Another popular strategy for zero-shot learning is meta-learning. Meta learning is also known as ``learning to learn'', which aims at training a meta learner which learns the update rules of the target model~\cite{GuWCLC18}. This enables a machine learning model to achieve competitive performance even with scarce data. However, meta-learning focuses on learning strategies instead of representations. Hence it will be difficult to be generalized across different code intelligence tasks. 
% Learning to discriminate is also a well-known technique for meta-learning. This category of techniques focuses on diving differences between samples unseen in training procedure and inferring convincing result~\cite{VinyalsBLKW16}. For example, the ``N-way-K-shot'' method \cite{WangYKN20} aims at learning how to classify few-shot examples. \textit{We use many groups of classification task samples with N labels for training. Each group contains K data samples as support set for training and a query set for evaluating. Finally this model can classify N labels task with only K data samples for training, even they did not show in the past training set.}

% For example, we can employ the GAN (generative adversarial networks) model which contains an encoder and a decoder to mutually promote each other with unsupervised data set~\cite{}. This enhance the model to adapt to different domains. \cn{N-way-K-shot, classical meta learning framework}

\textit{Pre-trained Language Models.} 
PLMs are pre-trained on large-scale text corpora to learn common knowledge of the languages, and can be generalized to specific tasks with only a few training examples. However, PLMs require a fine-tuning phase to adapt the pre-trained model to the downstream tasks, which needs the availability of labelled datasets. Therefore, it's still quite laborious to annotate the data manually.

\textit{Prompt-based Learning.} 
% \subsection{Prompt-based Learning} 
 To alleviate the data hungry problem of fine-tuning, \citet{gpt3} introduced prompt-based learning, a lightweight alternative of fine-tuning for PLMs. Unlike fine-tuning, which adds fine-tuning header and re-optimizes the PLM using downstream tasks, the prompt-based learning approach converts downstream tasks (e.g., method name detection) to the same form as the pre-training tasks (e.g., MLM) by injecting ``prompts'' and ``[MASK]'' to the PLM input. Hence, the PLM generates the target result after minimal adjustment. This encourages downstream tasks to reuse the knowledge from the PLM. 

\section{Approach}
\label{sec:approach}

\begin{figure*}[!htbp]
    \centerline{\includegraphics[width=0.6\textwidth]{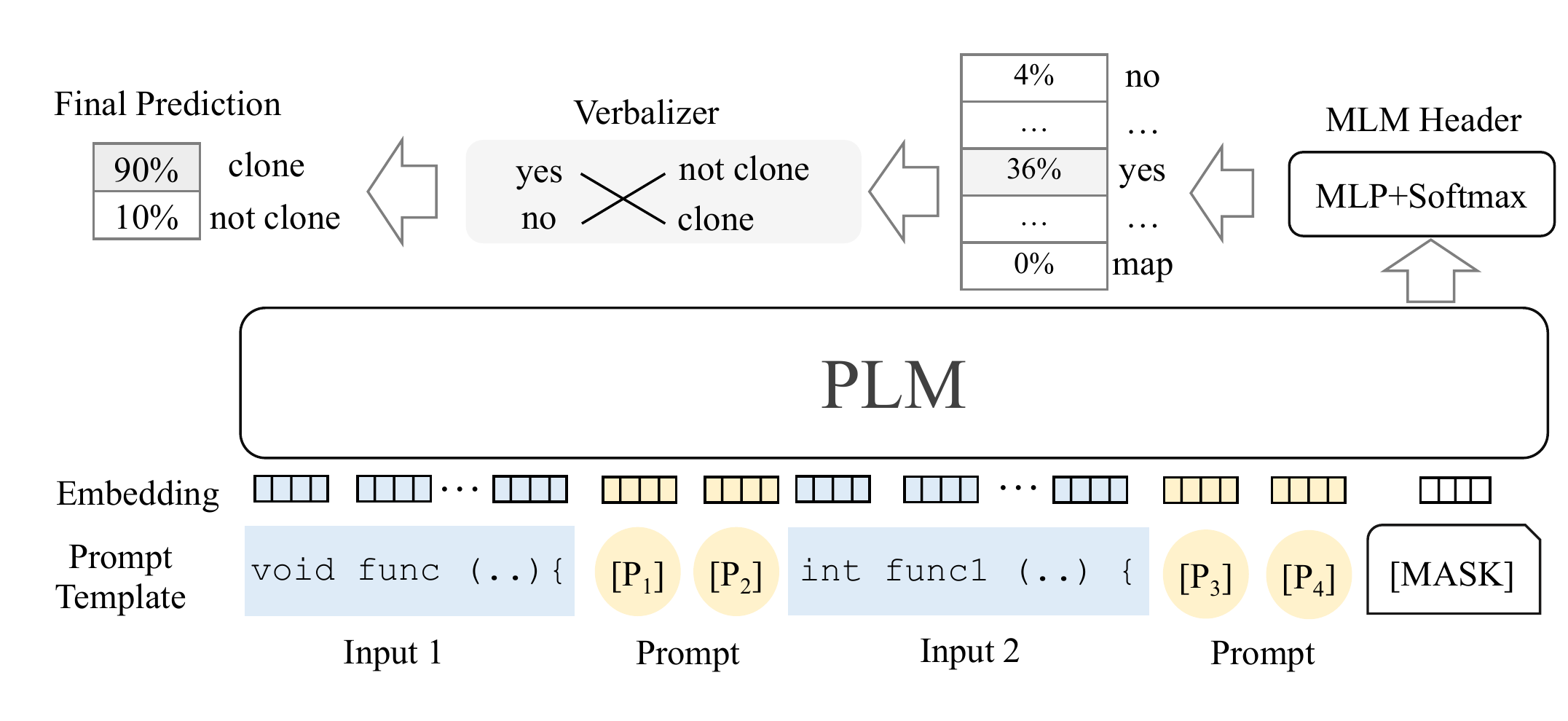}}
    \vspace{-2mm}  
    \caption{Model Architecture.}
    \vspace{-2mm}  
    \label{fig:structure}
\end{figure*}

\subsection{Problem Definition}
\textbf{Program Representation Learning:} Let $x=\{x_1,...,x_N\}\in\mathcal{X}$ denote a program code snippet with $N$ tokens. The goal of program representation learning is to map $x$ into an $d$-dimensional vector which contains program semantics, namely, $f_\theta:\mathcal{X}\rightarrow R^d$~\cite{BengioCV13}, 
%Program representations are code semantics in the format of vectors. \cn{Formula} They are unified and highly informative data format for representing code snippets. 
where $f_\theta$ is a function parameterized by $\theta$. $f_\theta$ can be implemented using deep neural networks such as the fully-connected networks~\cite{code2vec}, LSTM~\cite{MahtoVTH21} and Transformers~\cite{transformer}. 

The learned program vectors can further be taken as input to machine learning models for code intelligence tasks such as code clone detection and code search. Given two code snippets $x_1$ and $x_2$, we can train a classifier which predicts the class label~$y\in\mathcal{Y}$ given their vectors: 
\begin{equation}
\label{eq:task}
    p(y|x_1, x_2) = g_\phi(f_\theta(x_1), f_\theta(x_2)),
\end{equation}
where $g_\phi$ denotes the neural network model for the classification task.
For example, in the code clone detection task, $x_1$ and $x_2$ stand for two code snippets and $y$ stands for whether they contain clones. In the code search task, $x_1$ and $x_2$ stand for a code snippet and a natural language description, respectively, and $y$ stands for whether they are semantically correlated.

\smallskip\noindent\textbf{Zero-Shot Program Representation Learning:} Let $(\mathcal{X}^S, \mathcal{Y}^S)$ denote the labelled dataset in the source language, and $\mathcal{X}^T$ denote the unlabelled dataset in the target language. The goal of zero-shot program representation learning is to train a machine learning model which predicts the labels in the target language given the labelled data in the source language, namely, to estimate $p$, the probability of the label predicted:
\begin{equation}
    p(y^T| {x_1^T, x_2^T, \mathcal{X}^S, \mathcal{Y}^S)},
\end{equation}
where $x_1^T, x_2^T\in\mathcal{X}^T$ and $y^T\in\mathcal{Y}^T$. 
Representations for $\mathcal{X}^S$ might not be useful for $\mathcal{X}^T$ since there could be a lexical gap between $\mathcal{X}^S$ and $\mathcal{X}^T$. However, both of them can be used in pre-training tasks such as the MLM~\cite{codebert} for a PLM. Hence, it is feasible to bridge their representation gap using the common pre-training task. 
Based on this idea, we cast the problem in Equation~\ref{eq:task} as an MLM task for a PLM and train the model on ($\mathcal{X}^S$, $\mathcal{Y}^S$), so that the PLM can also predict the class labels for data in $\mathcal{X}^T$ seamlessly. 
%In zero-shot learning, the representation generating approach should have transfer ability for learning without large-scale code corpora.
%Because of this, good representations should demonstrate good interpretability and generalizability. In our work, ``zero-shot" focus on the situation where program languages have no code corpora for learning representations. Through our approach, we can get good representations of these zero-shot programs without their corpora.

\subsection{Model Architecture}

Domain-specific languages such as Solidity often have insufficient data for code intelligence tasks, while popular languages such as Java and Python have large-scale code corpora. Our goal is to transfer the representations of a popular programming language into a domain-specific language that has no training samples. As such, we keep pre-training the PLM on the task of a popular source language and then transfer the model to tasks in the target language.

Figure~\ref{fig:structure} illustrates the overall architecture of our \approach. 
The pipeline is comprised of three steps: 
%pre-training and prompt-based learning. We first pre-train a programming language model in a common language such as Java. The pre-trained model is then adapted into a downstream task in a target language through prompt-based learning. 
First, we cast any downstream task to the pre-training task (e.g., MLM) by inserting trainable prompts and a ``[MASK]'' token into the input of the task (\S 3.3). Taking the resulting data as input, a PLM is then re-trained using the MLM task, that is, infers the program representation of the input and predicts the word for the ``[MASK]'' token (\S 3.4). 
Finally, a verbalizer is employed to cast the predicted word to class labels (\S 3.5). 

\subsection{Casting Downstream Tasks into MLM}
\label{sec:prompt}
Our first step is to cast the downstream task (Equation~\ref{eq:task}) into the MLM task.
Given two input snippets $x_1$ and $x_2$ of the downstream task, we concatenate the two input snippets $x_1$ and $x_2$, inspired by NSP-BERT~\cite{nspbert}. %This input method can not only encode each segment semantic information, but also encode the relation information between them. 
%A sequence of segment ids~$s=s_1,...,s_{|x_1|+|x_2|}$ is also created to indicate whether each position $i$ belongs to $x_1$ ($s_i$=0) or $x_2$ ($s_i$=1). \cn{we did not use the segment ids}

% \gu{describe the MLM pre-training task.}
%Then, we aim to activate the pre-trained model to generate target labels. 
%We insert ``[mask]'' tokens into the original input and use the MLM header in the pre-training phase to search for the most likely word in the ``[mask]'' position. 
Like that in the MLM task, we also insert a ``[MASK]'' token into the concatenated input. The ``[MASK]'' token acts as a placeholder which steers the pre-trained model to generate the classification result~$y$ in the code intelligence task. 
It is notable that the position of the ``[MASK]'' token is a hyperparameter and we append it at the tail of the input by default. 
The masked sequence 
\begin{equation}
    \tilde{x} = [CLS];x_1;x_2;[MASK]
\end{equation}
is taken as input to the PLM, which yields the hidden states
\begin{equation}
    \mathbf{h}_1,...,\mathbf{h}_{|\tilde{x}|} = f_\theta(\tilde{x})
\end{equation}
for all tokens, where $f_\theta$ denotes the pre-trained programming language model parameterized by $\theta$.

Then, the hidden state corresponding to the masked token, namely, $\mathbf{h}_{-1}$, is fed into an MLM header $g_\phi$ which predicts a token for the masked position: 
\begin{equation}
    \hat{\mathbf y} = \mathrm{softmax}(g_\phi(\mathbf{h}_{-1})).
\end{equation}
The MLM header $g_\phi$ is a fully connected neural network parameterized by $\phi$ that is optimized to minimize the cross-entropy loss:
\begin{equation}
\label{eq:mlmloss}
    L_{\mathrm MLM}(\phi|\mathbf{y}, \hat{\mathbf y}) = - \sum_{i=1}^{|V|}y_i\mathrm{log}(\hat{y}_i),
\end{equation}
where $\mathbf{y}$ denotes the ground-truth label of the code intelligence task, and $|V|$ is the vocabulary size. 

\subsection{Prompt-based Learning}
The conventional fine-tuning method for the MLM task optimizes $f_\theta$ and $g_\phi$ in Equation~\ref{eq:task} from scratch. This causes the model to overfit scarce task-specific data. Inspired by prompt-based learning~\cite{ptuning}, we optimize the PLM by merely adjusting its input sequence. More specifically, we insert a number of pseudo tokens called prompt into the input sequence of the PLM, which coaxes the PLM to directly generate the predicted label of the downstream task. By only adjusting the model input, the PLM needs far less optimization cost to fit for the data in the target task, while keeping the most of prior knowledge learned during pre-training.

Based on this idea, we design a number of prompt tokens~$P=[P_1,...,P_m]$ and inject them into the masked sequence~$\tilde{x}$ using a pre-defined template~$T={[P]; x_1; [P]; x_2; [P]; [MASK]}$.
Hence, the original inputs $x_1$ and $x_2$ are transformed into
\begin{equation}
    \tilde{x} = [P_{1:i}]; x_1; [P_{i+1:j}]; x_2; [P_{j+1:m}];[MASK]
\end{equation}
through template $T$.  
For example, in the task of code clone detection, given two code snippets, ``code1'' and ``code2'', the transformed input is  
\begin{equation}
    \textsf{[$P_1$][$P_2$] \text{code1} [$P_3$] [$P_4$] \text{code2} [$P_5$][$P_6$][\text{MASK}]},
\end{equation}
which contains six trainable prompts in this case.

Like general words, these prompt tokens are embedded into trainable vectors and are optimized on downstream tasks in the target domain through gradient descend. %The template can be described as a transforming function $P$ and the input $\mathcal{X}$ after transforming is $P(\mathcal{X})$. %The template is also a hyperparameter and in \approach we enable 10 prompt tokens and evenly distribute them along with original token inputs and ``[MASK]".

%\cn{redundant with previous explanation of zero-shot learning?}
In a zero-shot setting, there is no training sample in a low-resource programming language. Instead, we train the PLM using large-scale code corpora in popular languages (e.g., Java) through the converted MLM task, and then directly apply the trained model to tasks in the low-resource language. 
More specifically, we train the PLM through prompt learning for the converted MLM task in the source domain. Then, we take as input data samples in the target domain into the same model without extra training, and obtain the results of the code intelligence tasks.

\subsection{Reverting MLM Outputs to Classification Labels}
\label{sec:veb}

The MLM task generates a token that is likely to fill into the masked position. In order to obtain the classification result, we need to revert the MLM predictions to classification labels of the downstream task. 
For this purpose, we employ a verbalizer~\cite{pet} which realizes such a reversion. Let $\mathcal{V}$ be the vocabulary of the PLM and $\mathcal{Y}$ be the labels of the downstream task such as $\{$true, false$\}$. The verbalizer is defined as a function $v$: $\mathcal{C} \rightarrow \mathcal{Y}$ that maps each candidate word in the vocabulary to a classification label. The choice of candidate words is arbitrary as long as they are sufficiently different. The model will be trained to map candidate words to true predictions. In our approach, we consider two candidate words $\{yes, no\}\in\mathcal{V}$ as a candidate set $\mathcal{C}$ and only inspect which word in $\mathcal{C}$ is more likely to fill into the ``[MASK]'' position through PLM predictions. If the word ``yes'' has a higher probability to fill in the masked position, the verbalizer will map it to the label ``true'' and hence output a positive prediction for this task.

Take code clone detection as an example. Given two code snippets, the model constructs an input sequence by injecting a number of prompt tokens into the snippets, followed by a ``[MASK]'' token. The constructed sequence is fed into the PLM to predict the label $Y$, where $Y\in\{$``cloned'', ``not cloned''$\}$. %We only check which word in the candidate set $\mathcal{C}$=$\{$``yes'', ``no''$\}$ is more likely to be filled into the ``[MASK]'' by the PLM. 
The MLM header of the PLM outputs the probability of each candidate word for the masked position. If the candidate word ``yes'' has a higher probability, the verbalizer will map it to the class label ``cloned'', yielding the final prediction as ``cloned''. 

%Verbalizer takes the responsibility to collect loss of each candidate words and predict token and find out the candidate with smallest loss. Then it maps the candidate word to the final answer of downstream tasks. Verbalizer is not a neural network and does not need to be trained. The ability of predict token for ``[MASK]'' is from PLM.

\subsection{Training and Usage}
\begin{figure}[!tbp]
    \centerline{\includegraphics[width=0.47\textwidth]{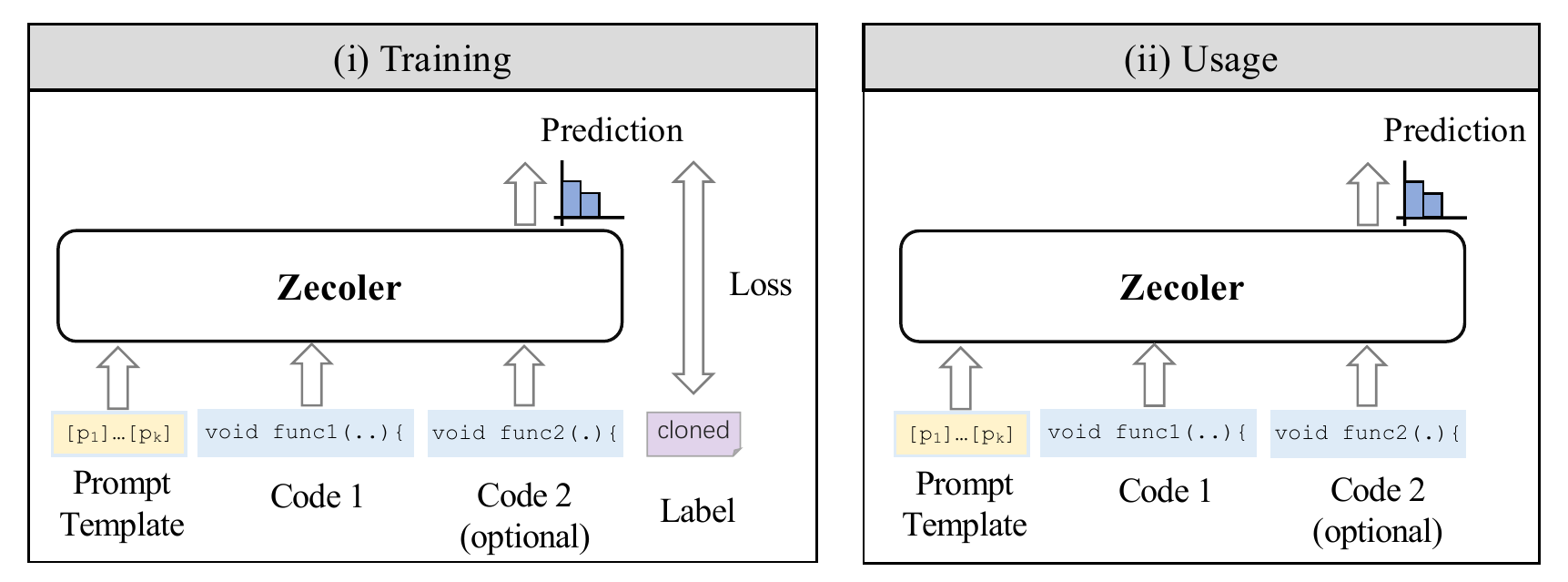}}
    \caption{The workflow of \approach. %First we pretrain the prompt tokens and PLM marked in yellow by training dataset, then we can deploy the model and get final answers based on MLM.
    }
    \label{fig:workflow}
\end{figure}

Figure~\ref{fig:workflow} shows the workflow of \approach. \approach follows the general paradigm of learning program representations. In the training phase, \approach is given a training set of labeled code snippets. For each snippet (pair), \approach augments it using a prompt template. The prompt-augmented code (pair) is taken as input to \approach which yields the prediction and calculates the loss function based on the ground-truth label. 

In the usage phase, \approach is given a code snippet (pair) only. \approach augments it using the same prompt template as in the training phase and then gives the prediction for the downstream task.

\section{Experimental Setup}
\label{sec:evaluation}

We implement our approach on top of CodeBERT~\cite{codebert}, one of the most popular pre-trained PLMs, and evaluate it on three code intelligence tasks. 

\subsection{Research Questions} 
The evaluation is designed to answer the following research questions:

    \begin{itemize}
        \item \textbf{RQ1: How effective is our approach in zero-shot program representation learning?}
        
        We evaluate the effectiveness of \approach in zero-shot program representation learning. We take Java as the source language and transfer the learned model to Solidity, a domain-specific languages, and Go, an up-and-coming language which are not provided with training samples. The experiments are conducted in three popular code intelligence tasks.
        
        \item \textbf{RQ2: How effective is our approach in few-shot program representation learning ?}
        
        In some programming languages and tasks, we can obtain small-scale training datasets. We wonder whether \approach is also effective for tasks with scarce (\eg, 100) data. Therefore, we provide the PLM models with a few samples of the target language and conduct the same experiments as in RQ1.
        
        \item \textbf{RQ3: How effective is our approach in monolingual program representation learning?}
        
        RQ1 and RQ2 mainly evaluate the effectiveness of \approach in a cross-language setting. We further explore how effective is our approach without transfer learning. We train the model in three languages, and test the model in the same language. Besides Solidity, we also want to assess our approach in other languages such as Java and Go when data labels are unavailable. 
        
        \item \textbf{RQ4: How do different hyperparameters impact the performance of our approach?}
        
        Finally, we evaluate the performance of our approach under different hyperparameters. Specifically, we conduct ablation studies on prompt templates (number and position), source languages, and PLM scales.
    \end{itemize}

\begin{table*}[!t]  
	\centering
	\caption{Overview of datasets.}
	\label{tab:dataset}
	\begin{threeparttable}
	\begin{tabular}{lccccc}  
		\toprule		      \multirow{2}{*}{Datasets} & \multicolumn{3}{c}{Downstream Tasks *} & \multirow{2}{*}{Size} & \multirow{2}{*}{Programming Languages} \\ \cline{2-4}
		\quad & CD & CS & MNP &   &   \\  
		\midrule   
		SCCD & $\surd$ & & & 10,000 & Solidity \\
		SCS~\cite{YangKYGWMZ21} & & $\surd$ & $\surd$ & 347,410 & Solidity \\
		CodeNet~\cite{codenet} & $\surd$ & $\surd$ &  & 8,008,527 & Java, Go, C++, C, Python, Ruby, C\#, ... \\
% 		BigCloneBench~\cite{bigclonebench} & $\surd$ & & & & \\
		CodeSearchNet~\cite{codesearchnet} & & & $\surd$ & 2,000,000 & Java, Go, Python, Javascript, Ruby, PHP \\
		\bottomrule  
	\end{tabular}
\begin{tablenotes}
\item * CD = clone detection, CS = code search, MNP = method name prediction.
%\vspace{-4mm}
\end{tablenotes}
\end{threeparttable}
\end{table*}

\subsection{Downstream Tasks} 
We evaluate our approach on three popular code intelligence tasks: 
%They are popular code intelligence tasks that have been widely used to evaluate program representations~\cite{}.
\begin{itemize}
\item[1)]\textbf{Code Clone Detection}: a task that determines whether two code snippets are cloned~\cite{bigclonebench} or not. A PLM based clone detection model takes as input two code snippets and outputs their representations. Then, a classification header is built on top of the representations and predicts whether the two code snippets are cloned (=1) or not (=0). There are four types of clones~\cite{clonedetection}. Our approach can challenge type-3 and type-4 clones, that is, the two snippets are not textually identical, but implement the same functionality.

\item[2)]\textbf{Code Search}: a task that retrieves a semantically relevant code snippet for a given natural language query~\cite{codesearchnet}. Following CodeBERT~\cite{codebert}, we formulate code search as a classification problem. Given a natural language description of the code and a programming language code snippet, this task aims at determining whether this NL-PL pair is related. The binary answer is ``related'' or ``not related''. The classification generates a probability score, which can be used for ranking results of code search.

\item[3)]\textbf{Method Name Prediction}: a task that suggests the function name for a given code snippet~\cite{ZhangCLP21}. Similar to code search, we transform this task to a binary classification task~\cite{ComptonFPK20}: given a code snippet, it enumerates all candidate function names (i.e., the vocabulary of code tokens) and constructs a ``<snippet, name>" pair. The pair is taken as input to the PLM which outputs a binary prediction whether the name in the pair is ``suitable'' (=1) or ``not suitable''(=0) for the code snippet.
\end{itemize}

%These three tasks are all with two input snippets and one output label. We show the adaptation way of \approach in Section \ref{sec:prompt} and Section \ref{sec:veb}. The way of baselines are described in Section \ref{sec:pretrain}.

\subsection{Datasets}

We conduct our experiments on four datasets. Each dataset is used for one or two tasks. Table~\ref{tab:dataset} shows the statistics of each dataset, including sizes, languages, and corresponding tasks.

\smallskip\noindent\textbf{Smart Contract Clone Detection (SCCD)}: a manually labelled clone detection dataset for the Solidity language. It contains 10,000 data samples collected from EtherScan, an analytic platform for smart contracts. We build web scrapers to collect Solidity code snippets of smart contracts and label the cloned pairs based on contract information such as contract address and opcode. Each data sample consists of a pair of code snippets that are cloned. One notable feature of this dataset is that most samples are type-3 and type-4 clones.

\smallskip\noindent\textbf{Smart Contract Summarization (SCS)}~\cite{YangKYGWMZ21}: a code summary dataset for the Solidity language. The dataset contains 347,410 code-comment pairs. It was originally collected for code summarization. We preprocess SCS to a classification format so as to fit for the code search and method name prediction tasks. We generate the code search dataset by filtering long code and removing code comments. We generate the method name prediction dataset by separating method name from original code snippets in SCS.

\smallskip\noindent\textbf{CodeNet}~\cite{codenet}: a multilingual codebase built from two online judge websites, namely, AIZU\footnote{https://onlinejudge.u-aizu.ac.jp/introduction} and AtCoder\footnote{https://atcoder.jp/}.  CodeNet contains 8,008,527 code submissions in multiple programming languages such as Java, Go, Ruby, and Python. We use this dataset for the code clone detection and code search tasks. To adapt the dataset to the code clone detection task, we label two code submissions as a cloned pair if they solve the same problem. %and have similar states (\eg, Accepted), time and space cost. 
To adapt the original data to the code search task, we extract NL-PL pairs from problem descriptions and their code submissions, respectively. 

% \textbf{BigCloneBench} (for clone detection):
% BigCloneBench is a wide used dataset for clone detection and built by Java. The Java clone detection dataset in BigCloneBench is different from that in CodeNet because of different problem domain. The problem domain means the topics which the problem covers. For example, we do not use a model for detecting human face to detect whether there is a bike in a picture, because these two tasks are in different problem domain.
 
\smallskip\noindent\textbf{CodeSearchNet}~\cite{codesearchnet}: a widely used code search dataset, which is used for the method name prediction task in our work. The dataset involves six languages, namely, Java, Go, Python, Javascript, Ruby and PHP, with 2,000,000 code snippets and corresponding method names.

We preprocess these datasets by filtering out comments. Code snippets with more than 250 tokens are filtered out to fit for the PLMs. We also exclude code snippets with less than 125 tokens to accommodate downstream tasks. 
In order to prevent the model from being biased to one class, we balance the dataset with the same number (1:1) of positive and negative samples. 
The negative pairs of code snippets ($y=0$) are created using random combinations of snippets from the positive data samples ($y=1$). %In detail, All the data samples of these data set are in the form of pair $\{(x_1^{(i)}, x_2^{(i)}, y^{(i)}) | x_1^{(i)}, x_2^{(i)}\in\mathcal{X}, y^{(i)}\in\mathcal{Y}\}$, in which $x_1^{(i)}, x_2^{(i)}$ denote two input snippets and $y_i^{(i)}$ means their label. We take the original labelled dataset as positive samples~$\{\mathcal{S}+, \mathcal{Y}+\}$, where $\mathcal{S}+$ contains all original snippet pairs, namely, $\mathcal{S}+ = \{(X_i, X_j) | X_i, X_j \in \mathcal{X}\}$ and $\mathcal{Y}+$ means the positive label. We rearrange two snippets in original data sample pairs and generate negative samples $\{\mathcal{S}-, \mathcal{Y}-\}$, where $\mathcal{X}- = \{(X_m, X_n) | (X_m, X_n) \notin \mathcal{S}+ \land X_m, X_n \in \mathcal{X}\}$ and $\mathcal{Y}-$ means negative label.

\subsection{Implementation Details}
We implement our models on top of the popular CodeBERT which is built based on RoBERTa-base (H=768, A=12, L=12). CodeBERT learns representations of programming languages (Java, Python, JavaScript, PHP, Ruby, and Go) in the pre-training phase. We use the default tokenizer (i.e., Microsoft/codebert-base) of CodeBERT with a vocabulary size of 50,265. We set the maximum sequence length to 512. 
Our experimental implementation is based on the Huggingface Transformers\footnote{https://huggingface.co/microsoft/codebert-base} and P-Tuning~\cite{ptuning}. The batch size and the number of epochs are set to 10 and 20 respectively. We insert prompt tokens to the original input of CodeBERT and place them uniformly.

All models are optimized using the AdamW~\cite{adamw} algorithm on a machine with two GeForce RTX 2080 Ti GPUs. The initial learning rate (lr) is set to 3e-5, 
which linearly increases from 0 during a warm-up period. The iteration number of the warm-up period equals to the number of the training step in first epoch. Then the learning rate decreases to 0 during the rest training process.
We measure the performance on the validation set during training. The checkpoint that achieves the best accuracy on the validation set are selected for testing. 

\subsection{Baseline Models}
\label{sec:baseline}

We compare our approach with five baseline models: %All of these baselines except NoPretrain are adapted to downstream tasks by fine-tuning method.
\begin{itemize}
\item[1)] \textbf{AVG}: a baseline approach that directly represents programs by averaging their token embeddings. We reuse token embeddings from CodeBERT and represent an input code snippet by the average of all token embeddings. Next, we fine-tune the classifier of downstream tasks using a 3-layer MLP header.

% \textbf{No-Pretrain}: \cn{directly use CodeBERT without any train? it does not make sense because the fine-tuning header is a MLP and will not work without any training. Or use MLM header instead?}~\gu{only MLM header}

\item[2)] \textbf{RoBERTa}~\cite{roberta}: a popular pre-trained language model that has also been used for programming languages~\cite{codebert}. The model is constructed with 12 transformer layers and pre-trained on a large English corpus with the MLM objective. We fine-tune it with a 3-layer MLP header over the ``[CLS]'' position.

\item[3)] \textbf{RoBERTa-large\footnote{https://huggingface.co/roberta-large}}:  
a large version of RoBERTa (H=1024, A=16, L=24)  with around 300 million parameters, which is almost twice the size of the normal version. We compare with this model to verify the advantages of \approach over large-scale PLMs.

\item[4)] \textbf{CodeBERTa\footnote{https://huggingface.co/huggingface/CodeBERTa-small-v1}}:
a version of RoBERTa pre-trained with CodeSearchNet, which was proposed by Huggingface. We use its default setting in our experiments.
%The number of layers is only half of the original RoBERTa configuration.

\item[5)] \textbf{CodeBERT}~\cite{codebert}: one of the state-of-the-art models for learning program representations. A more detailed description of CodeBERT can be found in Section~\ref{sec:pretrain}. We follow the same experimental setup in its original paper.
\end{itemize}

We implement these baseline models by referring to the work of CodeXGlue~\cite{codexglue}. We construct 3-layer fully connected neural networks as the fine-tuning header which maps the hidden vector of the ``[CLS]'' token to the class labels of downstream tasks. 

\section{Results}
%In this section, we analyze the experimental results to answer our research questions. 

\subsection{RQ1: Effectiveness of Zero-shot Learning}
In this experiment, we evaluate the effectiveness of \approach in zero-shot program representations learning. 
We initially train a representation model for each task using data samples of Java. Then, we adapt the trained model to the target languages (i.e., Solidity and Go) directly without extra training. We train the model with both 5,000 and 500 data samples of Java to assess the effects under different data sizes.

\begin{table}[!t]
	\centering
	\caption{Accuracy of program representation models on three tasks in the zero-shot setting.}
	\label{tab:rq1}
	\begin{threeparttable}
	\begin{tabular}{l@{}cccccc}
	\toprule
    \multirow{2}{*}{Model} & \multicolumn{2}{c}{CD} & \multicolumn{2}{c}{CS} & \multicolumn{2}{c}{MNP} \\ \cline{2-7} 
                              &Solidity&Go&Solidity&Go&Solidity&Go\\ \hline
    AVG               &   57.5    & 49.2      &  49.0     & 50.3     &  50.8     &  50.0  \\                          
    RoBERTa                      &   60.5    & 49.4      &  49.6     & 49.4      &  50.0     &  50.3  \\
    RoBERTa-L          &   47.3    & 51.0       & 48.7      & 48.8       &  51.7     &  48.5  \\
    CodeBERTa           &   57.9    &  67.3      &   53.2    & 53.1       &  49.7  & 49.0    \\
    CodeBERT                     &   65.4    & 91.7       &  48.9     & 46.2       &  52.1     &  65.2  \\\hline
    Zecoler\tiny{ 5000}       &   \textbf{79.8}    & \textbf{96.4}       &  \textbf{67.1}     & \textbf{80.3}        &  59.2     &  \textbf{98.8}   \\
    Zecoler\tiny{ 500}           &   74.9    & 82.4      &  53.3     & 56.9      &  \textbf{68.1}     &  90.4   \\ 
    \bottomrule
    \end{tabular}
\begin{tablenotes}
\item[*] The target languages (i.e., Solidity and Go) are not provided with training data. All source languages are trained with 5000 samples except the last one which is trained with only 500 samples.
\vspace{-4mm}
\end{tablenotes}    
    \end{threeparttable}
\end{table}

Table~\ref{tab:rq1} shows the accuracy of different models in three code intelligence tasks. We can observe that \approach significantly outperforms baseline models in all three tasks and all target languages. In the code clone detection task, the accuracy of \approach is 5\%-14\% greater than that of CodeBERT, the strongest baseline. The improvement is much more significant in the code search (30\% in average) and method name prediction (24\% in average) tasks. By contrast, AVG and RoBERTa-large obtain results that are close to random, indicating that they can hardly learn useful knowledge from few data samples.

% In particular, \approach shows outstanding performance in zero-shot learning. 
% For example, the result of Go in the clone detection task achieves a comparable accuracy that is only 2\% lower than that of the results in the source language (i.e., Java in RQ3). 
% In most of the cases, the accuracy of \approach in the zero-shot setting is higher than 70\%. 

The same trend can be observed when only 500 (1/10) samples of the source language are provided for training.
As the data size decreases from 5000 to 500, the accuracy of \approach drops in all tasks. Nevertheless, it still significantly outperforms the baseline models. % except the result of Java and Go in code clone detection. 
This means that \approach can learn representations much more efficiently while requiring smaller data compared with baselines.

It is notable that CodeBERT outperforms RoBERTa and CodeBERTa in both the code clone detection and method name prediction tasks, except for the code search task. We conjecture that CodeBERT is pre-trained on programming languages whereas the RoBERTa is only pre-trained on natural languages. Hence, CodeBERT can be better adapted to PL related tasks.
%RoBERTa can be better adapted to NL related tasks.
%\textit{Although we fine-tune both PLMs in Java for downstream tasks, the inputs of these tasks are mainly program languages and the ability gap formed in the pre-training procedure is hard to be closed.}

%\textit{Also, we can see that CodeBERTa outperforms RoBERTa with only half number of layers. This is because that CodeBERTa is pre-trained with CodeSearchNet and learn the knowledge of program languages in the pre-training procedure, which makes CodeBERTa more suitable for these code intelligence tasks. But it is still not as good as CodeBERT due to the lack of RTD training objective compared with the latter.}

%\cn{May Delete?} Another interesting observation is that \approach trained with 500 data samples outperforms that with 5,000 data samples in the method name prediction task for Solidity. One potential reason is that the model overfits the Java data in the training process, which results in inferior performance in the target low-resource languages such as Solidity.

%\textit{We can also find that RoBERTa-large performs poorly even containing double number of layers than that of RoBERTa. This means that too large PLMs may lose the ability of inferring when lacking training data of downstream tasks for fine-tuning. The overfitting problem in this case is serious.}

%Overall, the results suggest that \approach is effective in zero-shot learning of program representations in domain-specific languages such as Solidity and Go.

%\begin{framed}\noindent
\emph{Answer to RQ1:}
Our approach shows greater performance than baseline models in code intelligence tasks for no-resource programming languages, affirming the strong ability of \approach in zero-shot program representation learning.
%\end{framed}

% We can learn representation with the help of MLM in the PLM, which is powerful and born with the ability of generalization capability. MLM is also completely useful than fine-tuning when lacking training data. The trainable prompt makes the use of MLM even more suitable for our downstream tasks. Though the prompt trained can not be understood by human, it can guide the model to infer most likely result.

\subsection{RQ2: Effectiveness of Few-Shot Learning}
In this experiment, we evaluate the effectiveness of \approach in few-shot learning of program representations.
We continue training the model in RQ1 using a few data samples of the target languages. We vary the data sizes from 32 to 700 and evaluate the performance in three code intelligence tasks.
%For example, for Solidity clone detection with \approach, we use 100 extra data to continue training the model learned in RQ1 for zero-shot learning representation. 

\begin{figure}[tb]
    \centering
    %\subfigure[Result in Java]{
    %    \includegraphics[scale = 0.17, trim=10 10 10 10]{pics/tab3/cd_java_fs.png}
    %}
    \subfigure[Clone Detection (Solidity)]{
        \includegraphics[scale = 0.16, trim=10 10 10 10]{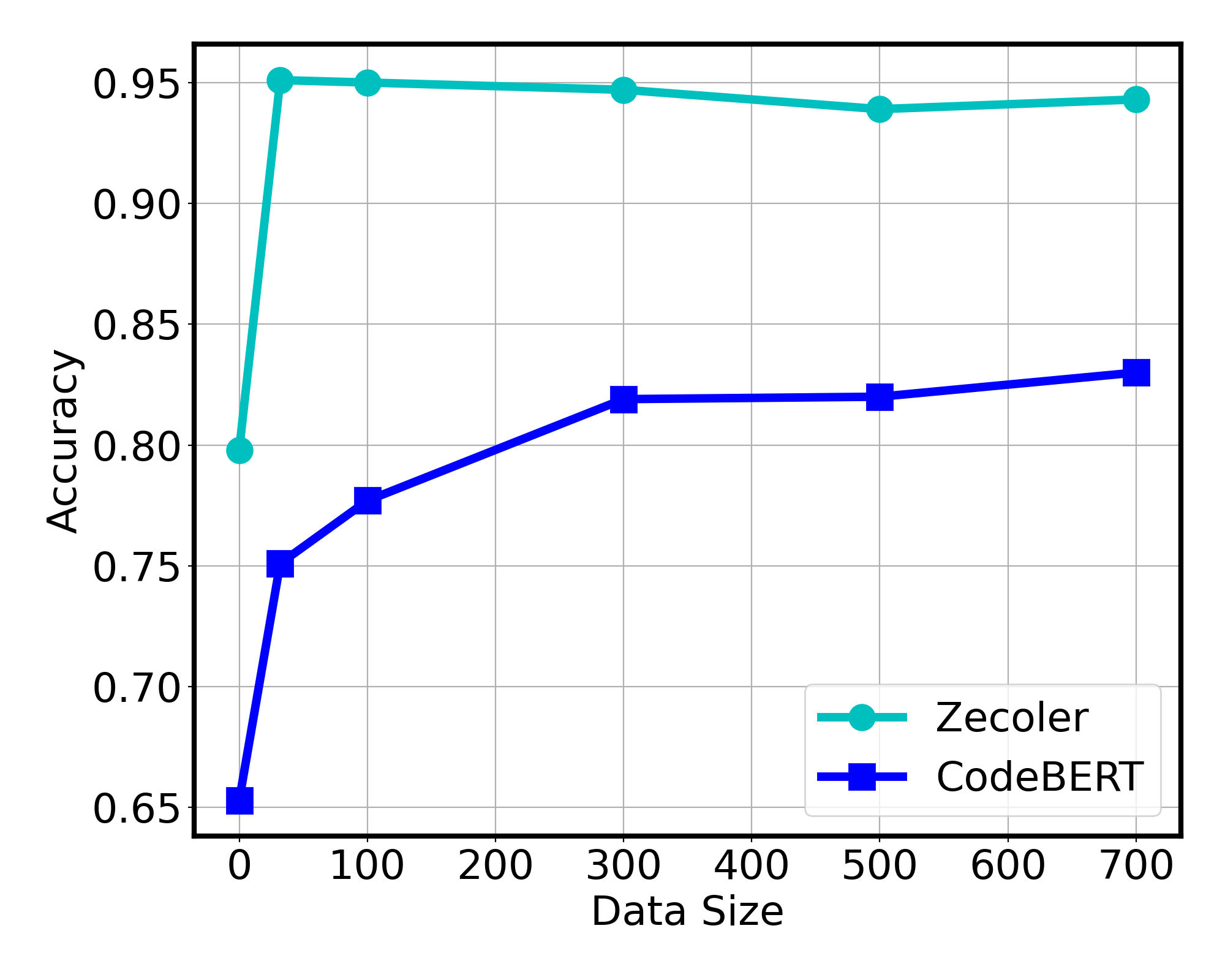}
    }
    \subfigure[Clone Detection (Go)]{
        \includegraphics[scale = 0.16, trim=10 10 10 10]{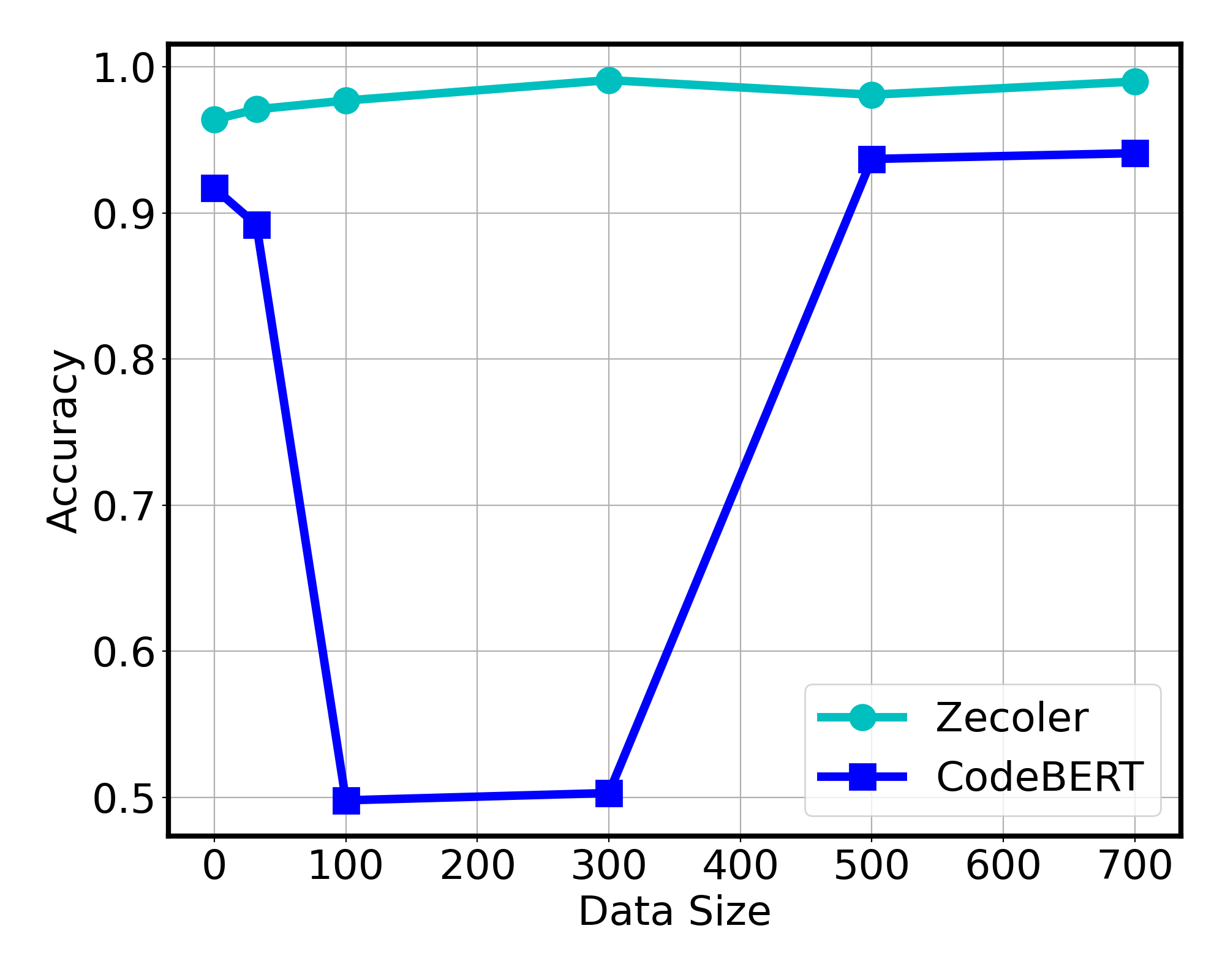}
    }\\
    \vspace{-6pt}
    %\subfigure[Result in Java]{
    %    \includegraphics[scale = 0.17, trim=10 10 10 10]{pics/tab3/cs_java_fs.png}
    %}
    \subfigure[Code Search (Solidity)]{
        \includegraphics[scale = 0.16, trim=10 10 10 10]{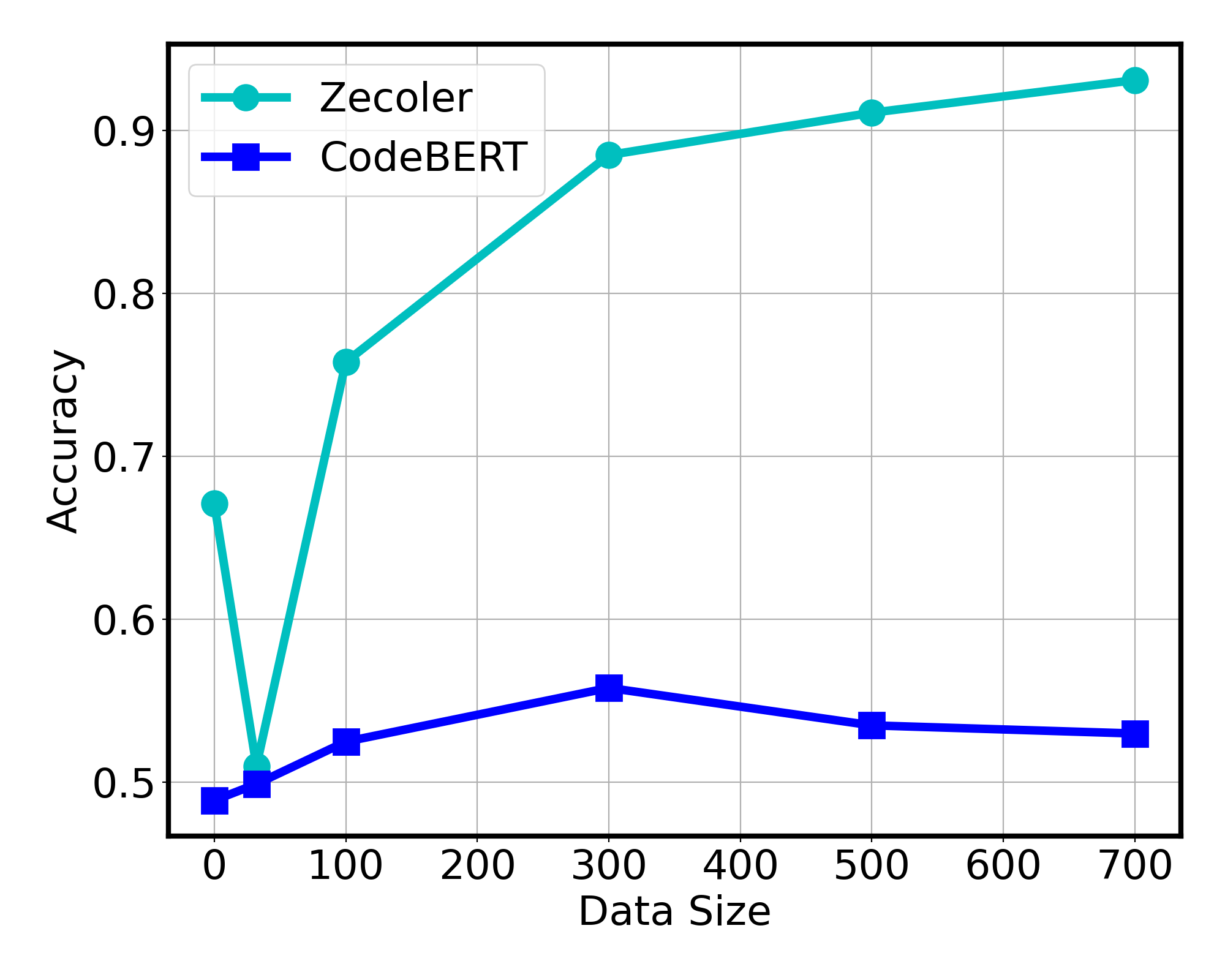}
    }
    \subfigure[Code Search (Go)]{
        \includegraphics[scale = 0.16, trim=10 10 10 10]{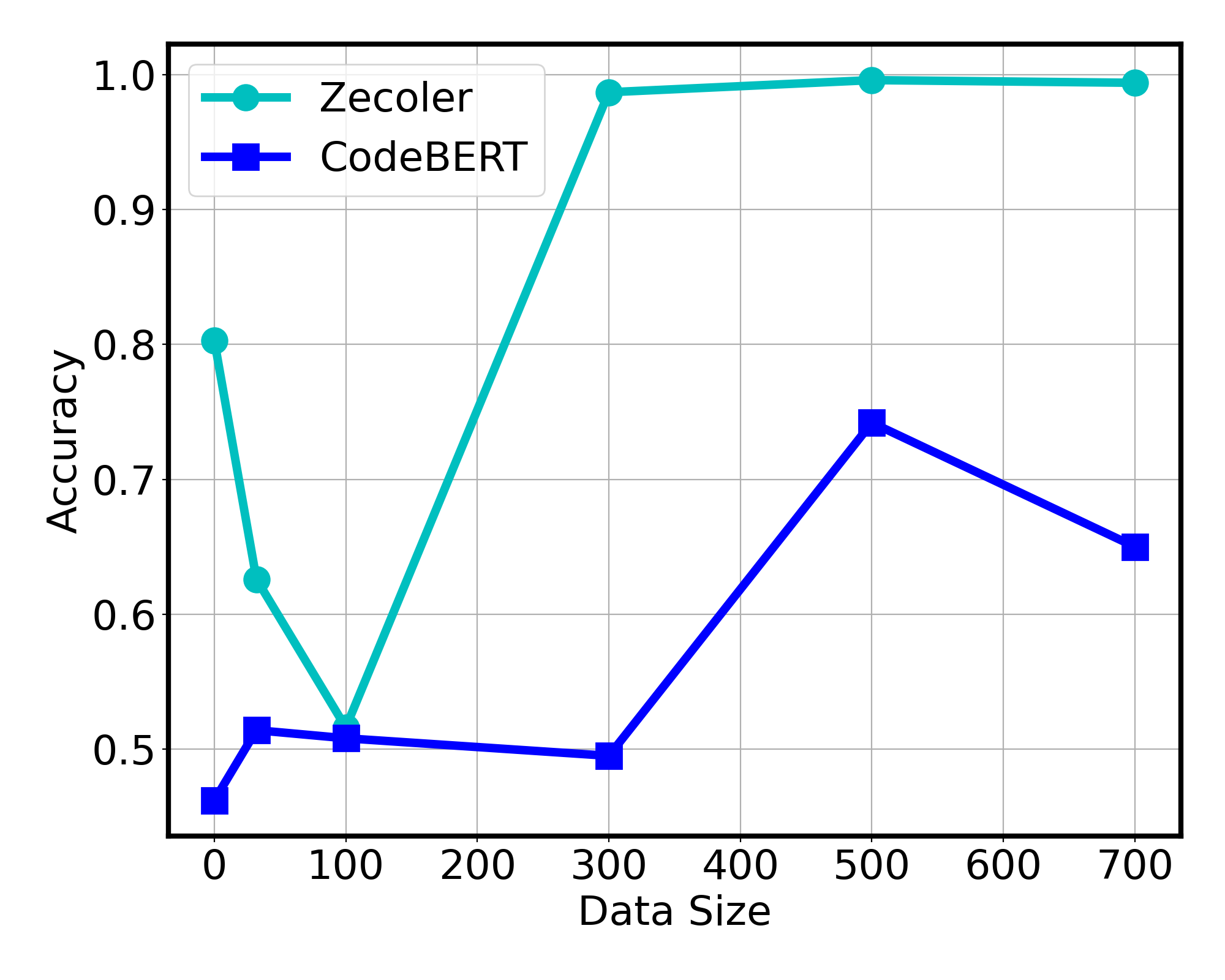}
    }\\
    \vspace{-6pt}
    %\subfigure[Result in Java]{
    %    \includegraphics[scale = 0.17, trim=10 10 10 10]{pics/tab3/mnp_java_fs.png}
    %}
    \subfigure[Method Name Prediction (Solidity)]{
        \includegraphics[scale = 0.16, trim=10 10 10 10]{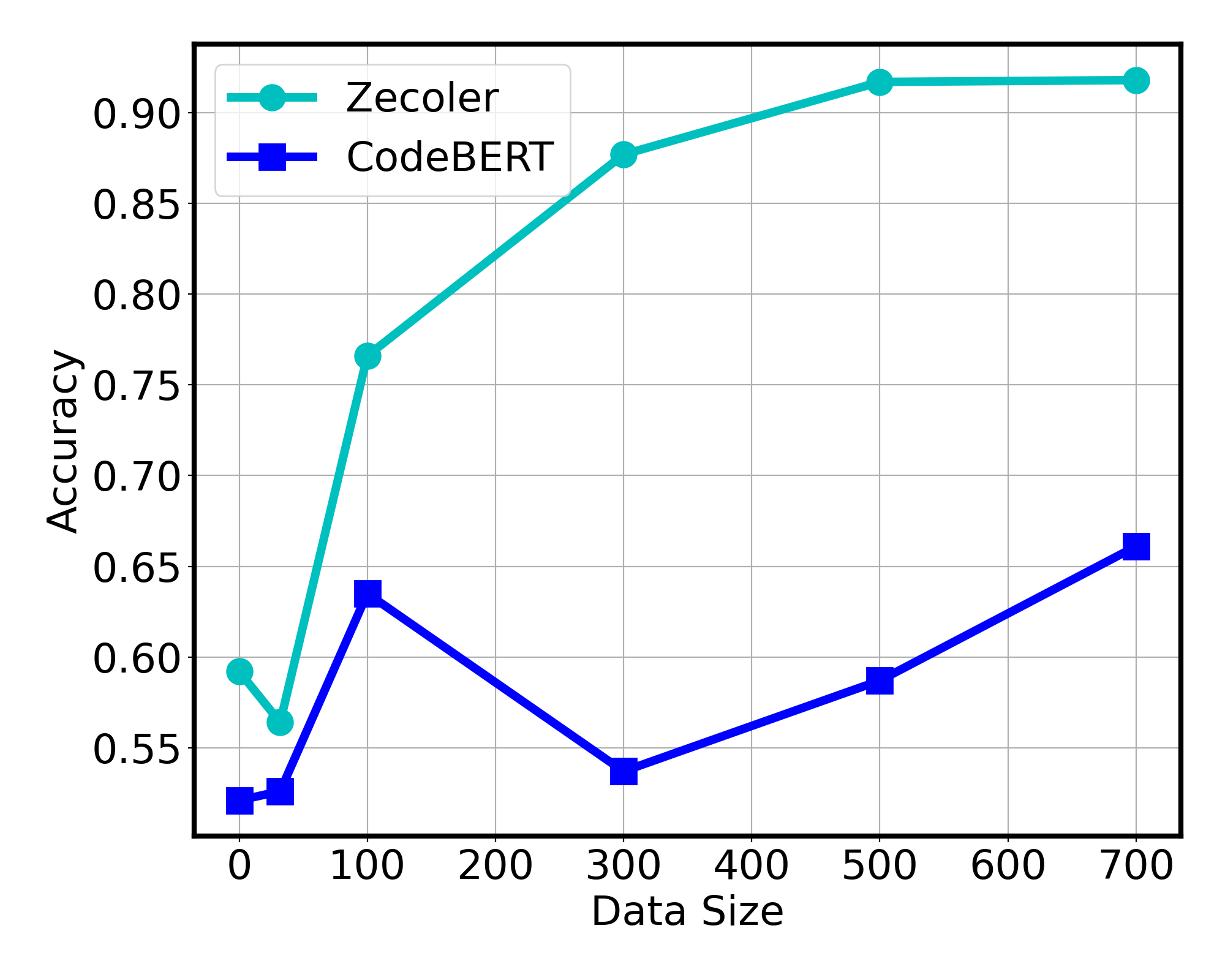}
    }
    \subfigure[Method Name Prediction (Go)]{
        \includegraphics[scale = 0.16, trim=10 10 10 10]{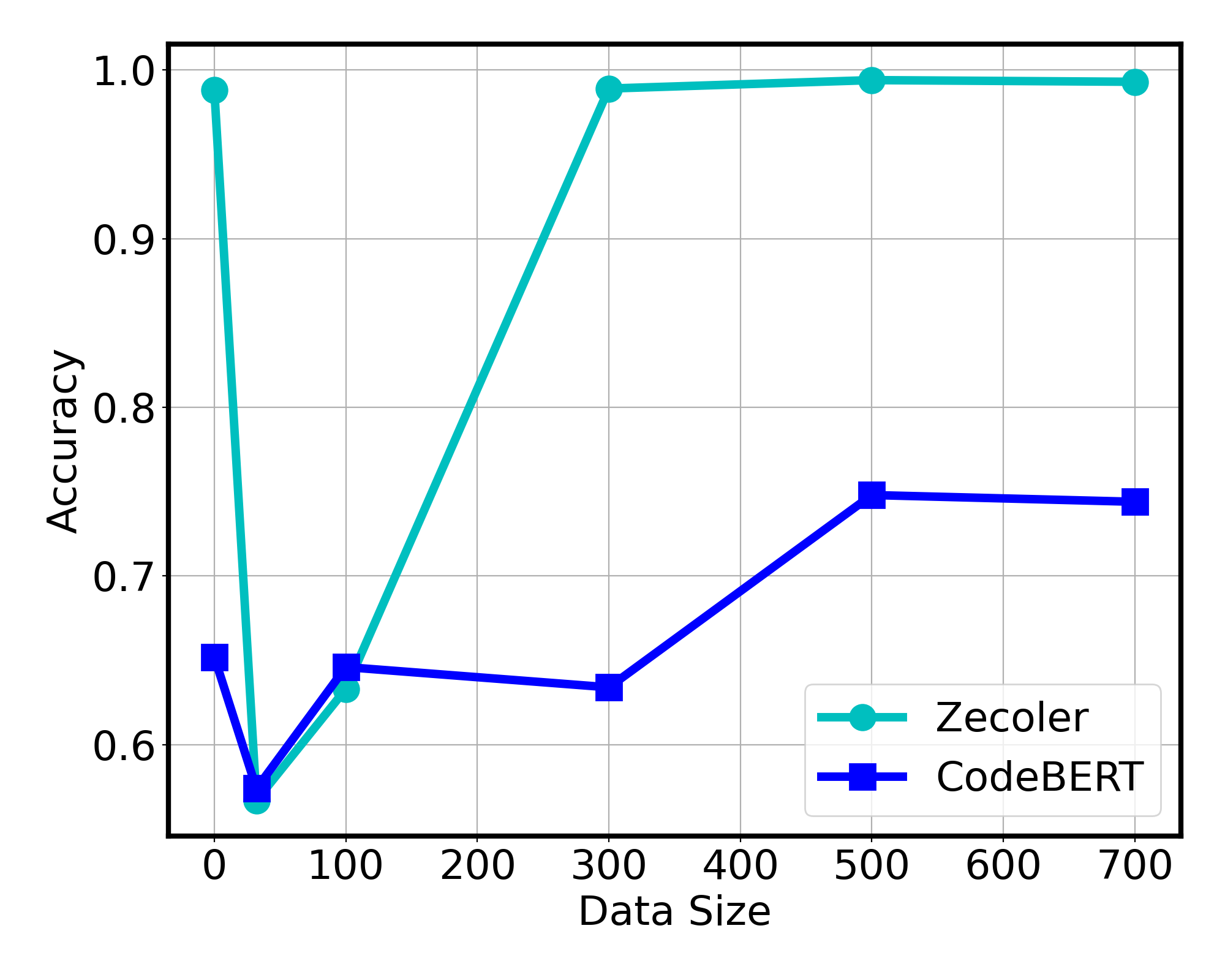}
    }
    \caption{Accuracy of program representation models continuously trained with different-scale datasets in few-shot setting.
    %Performance of \approach when feeding auxiliary dataset. %a-b: code clone detection, c-d: code search, e-f: method name prediction.
    }
    \label{fig:rq2}
\end{figure}

% have improvement, zecoler better
Figure~\ref{fig:rq2} shows the results. Compared to CodeBERT, the strongest baseline in RQ1, \approach shows greater strength in all tasks when provided with a few data samples of the target language. When the data size is 700, the accuracy of \approach is about 30\% greater than that of CodeBERT. %, and is higher than that of RQ1 in most of the experiments. 
This means that \approach can learn code representations more effectively when a small number of samples are available in the training dataset.
%in a few-shot setting.
% better than train from head
%Because of training in source program language in RQ1, the model already contains knowledge of downstream tasks. It is easier to get good performance by continuing training on these models instead of training on the original PLMs from the beginning. This is especially noticeable for CodeBERT that it archives acceptable results with 300 or 500 data size in this way, but at least 700 data size by training on the original PLMs showing in RQ3. 
We notice that when the data size is too small (e.g., 32), the model tends to overfit data. In this situation, zero-shot learning is preferred.
%In some cases, the small data size such as 32 may lead to very poor performance. This may be caused by overfitting since keeping training on such small-scale dataset reduces the ability of generalization. Reducing training step or multi-task learning may alleviate this problem.

%\begin{framed}\noindent
\emph{Answer to RQ2:}
\approach shows effective performance with a few data samples. 
As in zero-shot setting, \approach still keeps the best accuracy among all compared models in few-shot setting. 
%\cn{\approach performs better than baseline model with a few data samples. This means \approach still works in few-shot scenario.}
%\end{framed}

\begin{table*}[!t]  
	\centering
	\caption{Accuracy of various program representation models in a monolingual few-shot setting.}
	\label{tab:rq3}
	\begin{threeparttable}
	\begin{tabular}{lcccccccccc}
	\toprule
    \multirow{2}{*}{Model} & \multicolumn{3}{c}{CD} & \multicolumn{3}{c}{CS} & \multicolumn{3}{c}{MNP} & \multirow{2}{*}{Average} \\ \cline{2-10} 
                              & Java     & Solidity & Go     & Java   & Solidity  & Go     & Java    & Solidity  & Go   &   \\ \hline
    AVG\tiny{ 300}               &  51.6    & 64.1     & 50.1   & 49.1   & 50.2      & 50.1   & 47.9    &  48.1     & 50.1 & 51.3  \\
    RoBERTa\tiny{ 300}                    &  46.6    & 73.5     & 50.1   & 52.5   & 55.2      & 50.1   & 50.1    &  53.7     & 50.1 & 53.5  \\
    RoBERTa-large\tiny{ 300}              &  53.0    & 75.9     & 53.9   & 50.4   & 57.4      & 51.4   & 47.8    &  55.6     & 50.1 & 55.1 \\
    CodeBERTa\tiny{ 300}    & 50.7    & 68.3     & 65.3  &  52.0  & 58.8     & 45.0   & 50.8  &  61.8     & 47.4  & 55.6 \\
    CodeBERT\tiny{ 300}                   &  51.3    & 69.4     & 49.5   & 50.8   & 56.5      & 49.5   & 49.3    &  53.9     & 49.5 & 53.3  \\\hline
    Zecoler\tiny{ 300}        &  \textbf{85.8}    & \textbf{94.3}     & 99.3   & 51.7   & \textbf{90.1}      & \textbf{99.5}   & \textbf{98.7}    &  \textbf{88.8}     & \textbf{99.2} & \textbf{89.7}  \\
    Zecoler\tiny{ 100}        &  63.6    & 93.9     & \textbf{99.5}   & \textbf{52.6}   & 63.0      & 95.7   & 72.8    &  62.8     & 77.7  & 75.7 \\ 
    \bottomrule
    \end{tabular}
\begin{tablenotes}
\item * In this experiment, we train and test the model in the same programming language respectively. The training data size of source languages is 300 except the last one with only 100 data samples.
%\vspace{-4mm}
\end{tablenotes}    
\end{threeparttable}
\end{table*}

\subsection{RQ3: Effectiveness of Monolingual Few-Shot Learning}

Different from RQ2 in a cross-language few-shot setting, in this experiment we evaluate the effectiveness of our approach in a monolingual few-shot setting. We train models with a few samples of Java, Solidity and Go, and evaluate the performance of the tasks in the same language. 

Table~\ref{tab:rq3} shows the accuracy of different approaches in three downstream tasks. 
%We use 300 data samples to continuously train the basic models for baselines, and vary the number of data samples from 100 and 300 for \approach. 
We can observe that \approach outperforms baselines in the monolingual few-shot setting. 
Most of the baseline models just predict random answers, with an accuracy of around 50\%. This indicates that the baseline models cannot learn meaningful program representations with scarce data. Comparatively, \approach achieves 75.7\% accuracy in average with only 100 data samples. The results suggest that \approach learns program representations efficiently in the monolingual few-shot setting.

\begin{figure}[tb]
    \centering
    %\subfigure[Result in Java]{
    %    \includegraphics[scale = 0.17, trim=10 10 10 10]{pics/tab3/cd_java_fs.png}
    %}
    \subfigure[Clone Detection (Solidity)]{
        \includegraphics[scale = 0.16, trim=10 10 10 10]{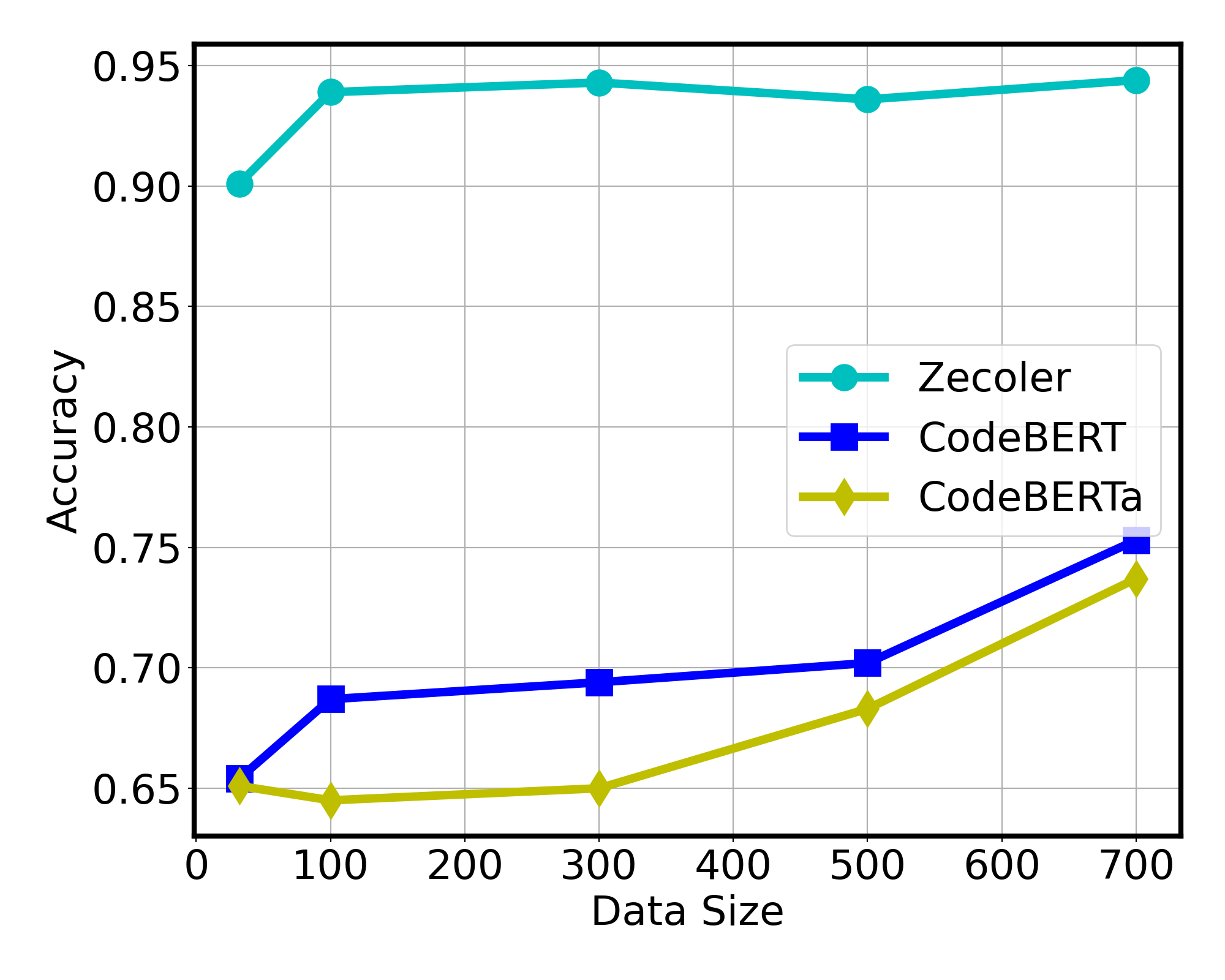}
    }
    \subfigure[Clone Detection (Go)]{
        \includegraphics[scale = 0.16, trim=10 10 10 10]{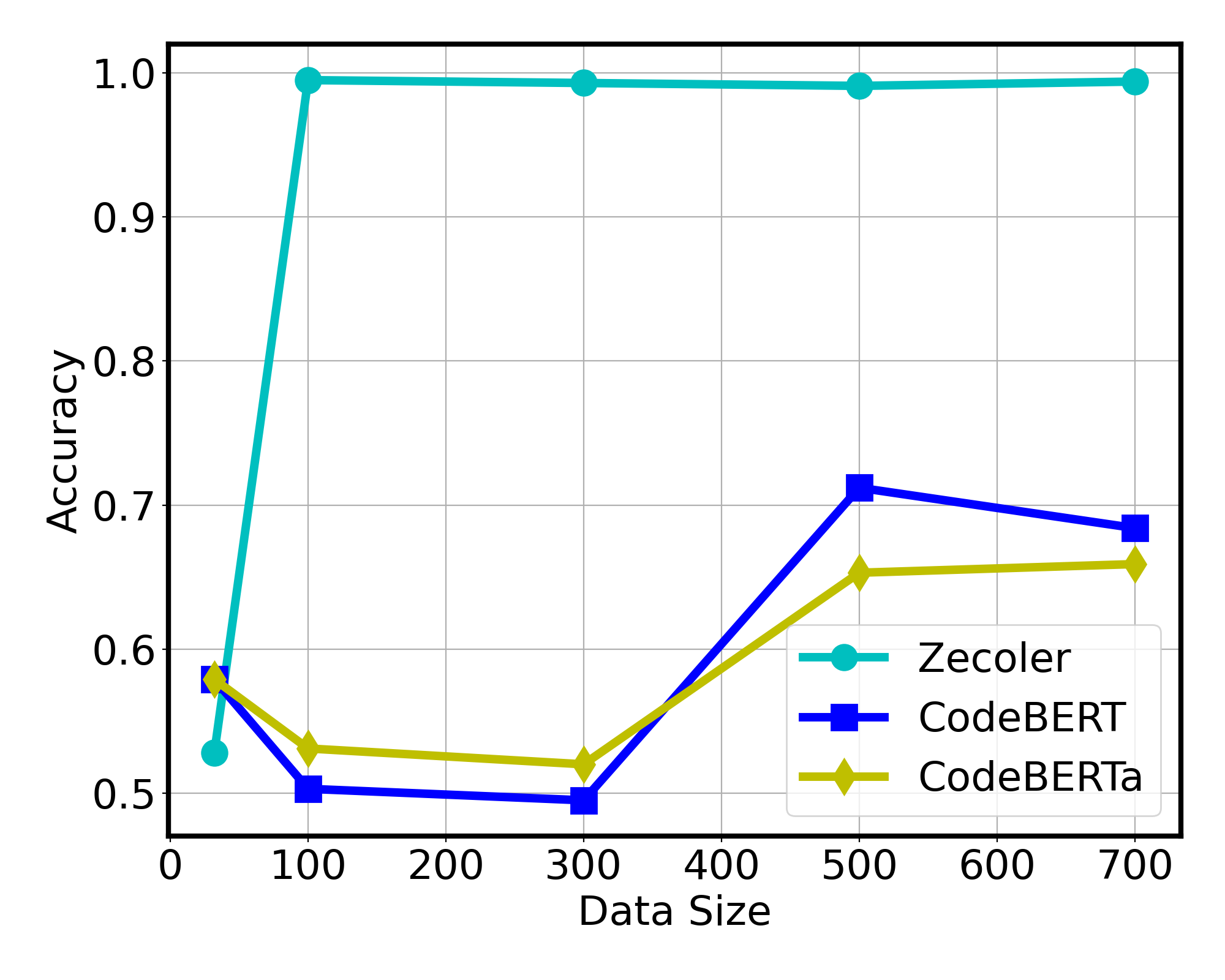}
    }\\
    \vspace{-6pt}
    %\subfigure[Result in Java]{
    %    \includegraphics[scale = 0.17, trim=10 10 10 10]{pics/tab3/cs_java_fs.png}
    %}
    \subfigure[Code Search (Solidity)]{
        \includegraphics[scale = 0.16, trim=10 10 10 10]{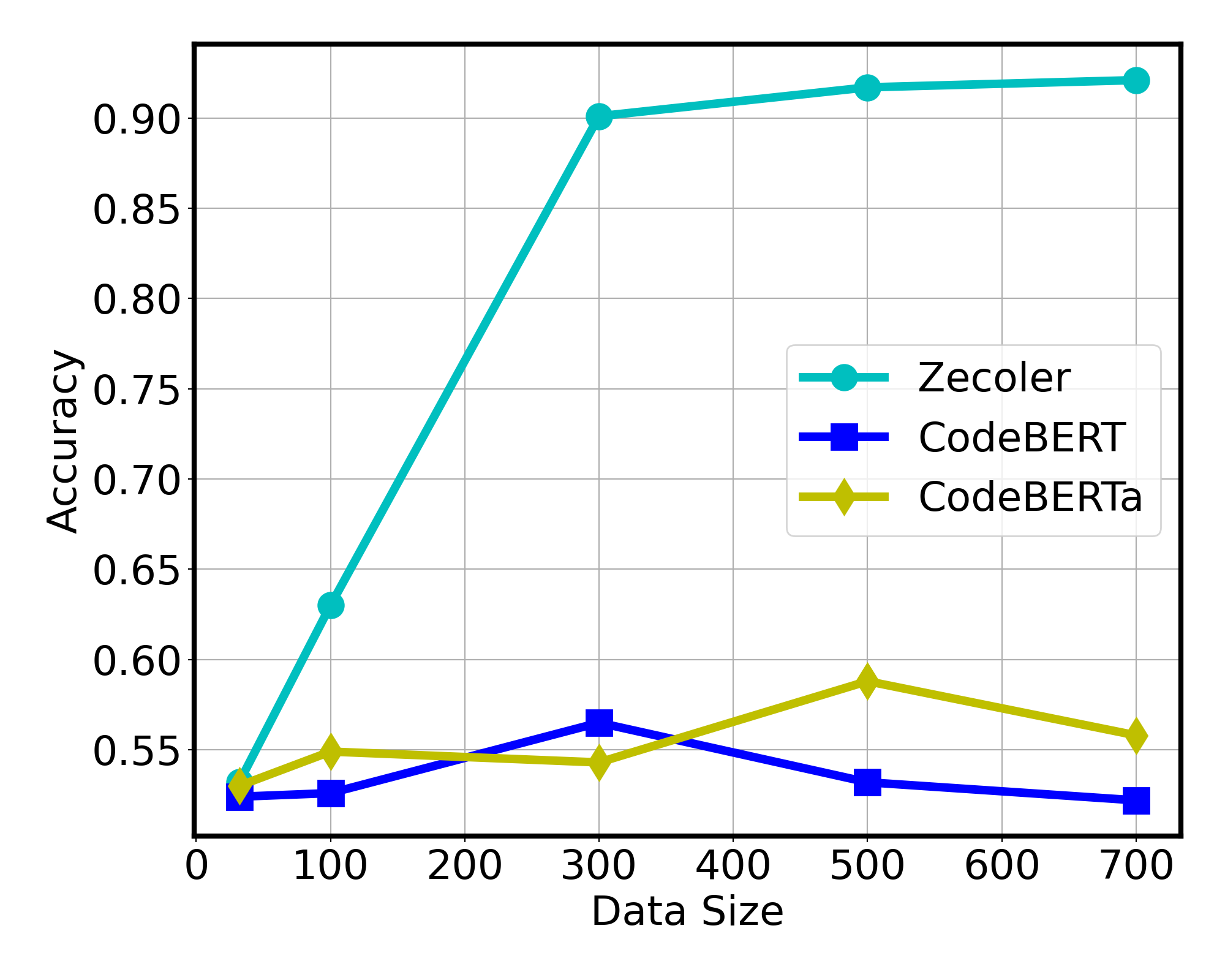}
    }
    \subfigure[Code Search (Go)]{
        \includegraphics[scale = 0.16, trim=10 10 10 10]{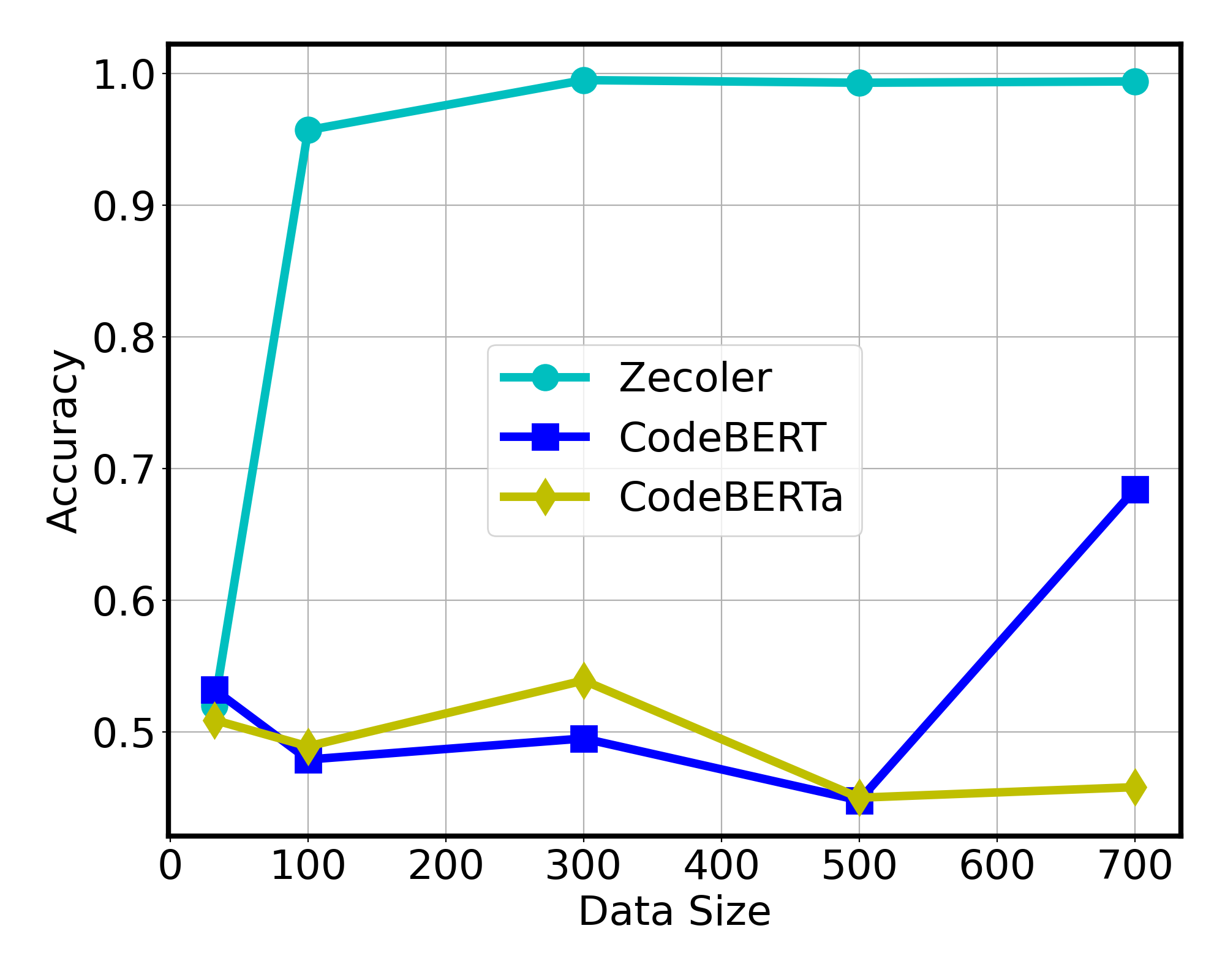}
    }\\
    \vspace{-6pt}
    %\subfigure[Result in Java]{
    %    \includegraphics[scale = 0.17, trim=10 10 10 10]{pics/tab3/mnp_java_fs.png}
    %}
    \subfigure[Method Name Prediction (Solidity)]{
        \includegraphics[scale = 0.16, trim=10 10 10 10]{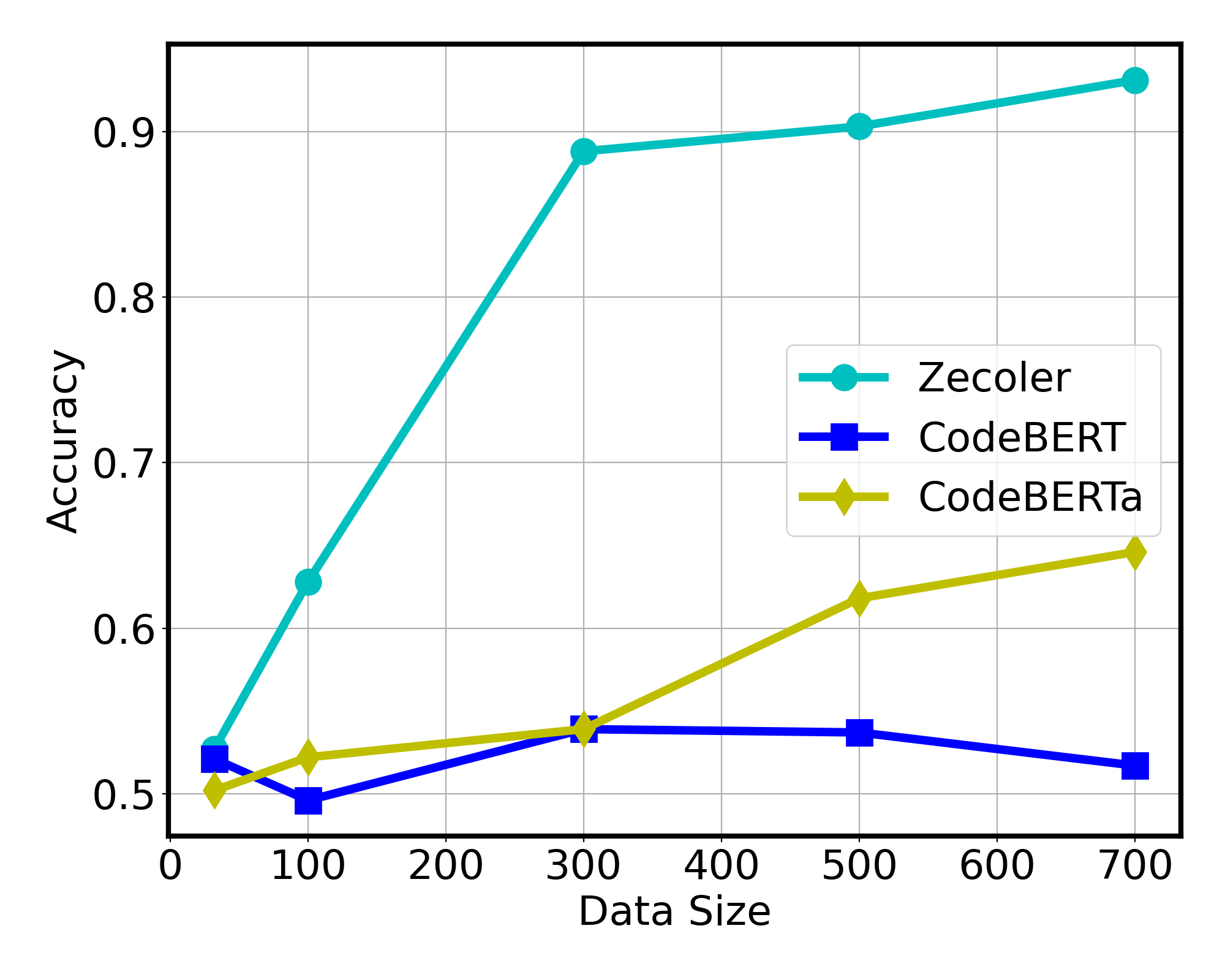}
    }
    \subfigure[Method Name Prediction (Go)]{
        \includegraphics[scale = 0.16, trim=10 10 10 10]{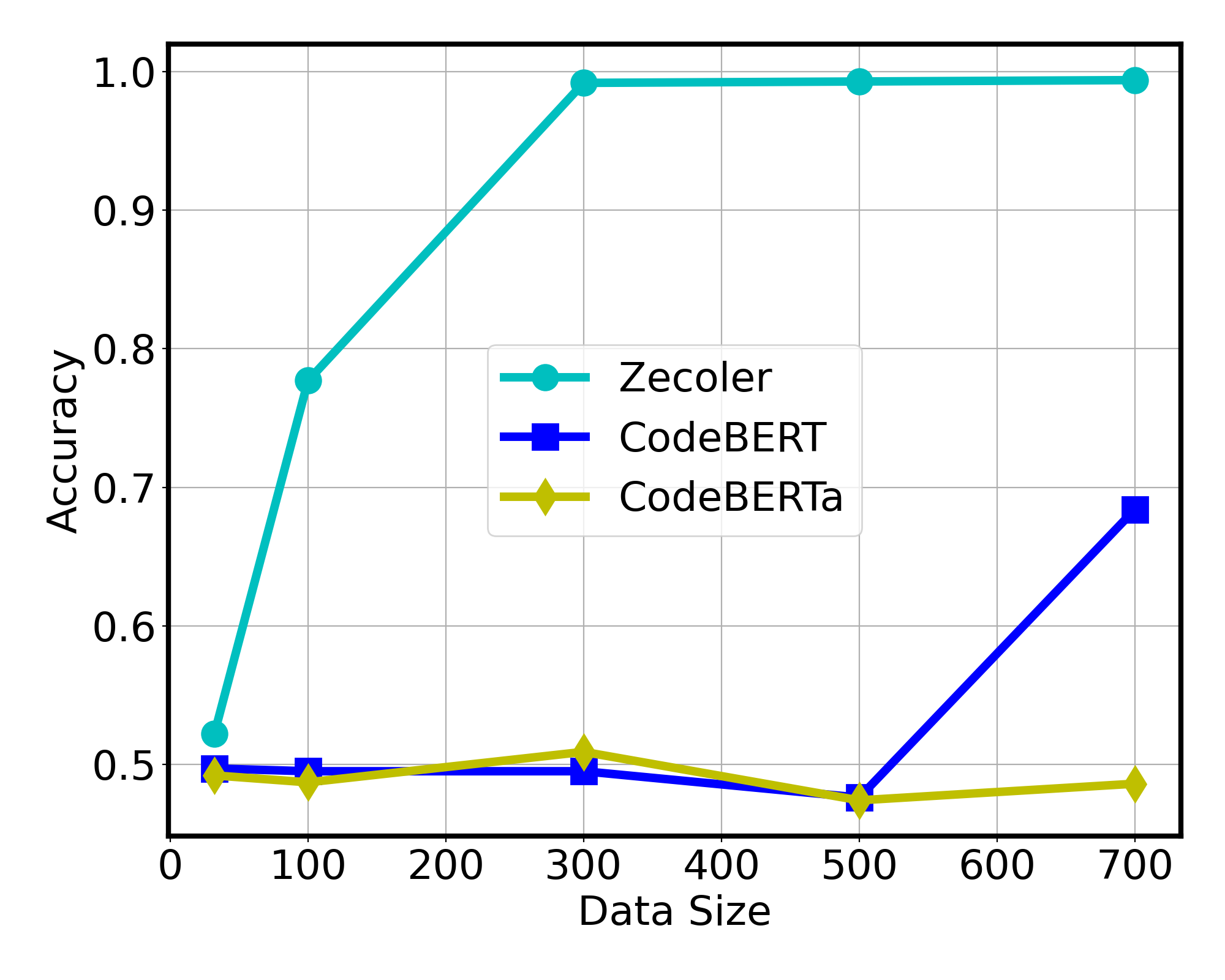}
    }
    \caption{Accuracy of program representation models continuously trained with different-scale datasets in monolingual few-shot setting.
    %Performance of \approach under different data sizes in three tasks. %(a-b: code clone detection, c-d: code search, e-f: method name prediction.)
    }
    \label{fig:rq3}
\end{figure}

Figure~\ref{fig:rq3} shows the performance of \approach, CodeBERT, and CodeBERTa with different data sizes in the code clone detection task. 
We can see that \approach outperforms the other two baselines under almost all data sizes. Furthermore, as the data size increases, the accuracy of \approach grows faster than that of baseline models. This indicates that \approach is effective in learning program representations given only 100 or 300 data samples. 

We have also observed that monolingual learning outperforms cross-language learning on small data sizes (\eg, 32 and 100), but achieves similar performance when the data size becomes larger. This is because continuously training on scarce data of a different language can lead to overfitting. 
% \cn{in most situation we do not have dataset in target program language.}
%In the case where we have small scale of dataset, it is better for us to use zero-shot learning or monolingual few-shot learning  instead of cross-language learning.

%There are also anomalies in Solidity of CD and Java of CS that \approach takes less data for training shows better results. These anomalies may be caused by the randomness of experiment and the gap is less than 1\%. 

%\begin{framed}\noindent
\emph{Answer to RQ3:}
\approach is effective in monolingual few-shot learning, and shows much stronger performance than that in the cross-language setting.
%especially in the case when we have extremely small scale of dataset compared with cross-language few-shot learning.
%meaning that the model learns good program representation when we only have a little training data of downstream tasks. 
%\end{framed}

\subsection{RQ4: Ablation Study}
\label{sec:ablation}
% Finally, we evaluate the performance of our approach in different settings of hyperparameters. Specifically, we conduct ablation studies of prompt templates (number and position), source languages, and PLM scales.
In this experiment, we inspect the performance of \approach under different hyperparameters. We vary the prompt templates and numbers of prompt tokens to search for the optimal prompt template. We also explore the impact to the performance by different source languages and different scales of backbone PLMs.

\smallskip\noindent\textbf{Prompt Templates:}
We first explore the effect of prompt templates on performance. We vary the position of the prompt tokens $P_{1:k}$ in the prompt template, namely,
head: [$P_{1:k}$, $x_1$, $x_2$, MASK], 
middle: [$x_1$, $P_{1:k}$, $x_2$, MASK], uniformly: [$P_{1:m}$, $x_1$,$P_{m+1:n}$, $x_2$,$P_{n+1:k}$, MASK] and
tail: [$x_1$, $x_2$, MASK, $P_{1:k}$].
% as Figure \ref{fig:prompt_position} shows. \cn{}.
% \begin{figure}[tb]
%     \centerline{\includegraphics[width=0.5\textwidth, trim=0 80 0 50,clip]{pics/prompt_position.pdf}}
%     \caption{The different position of prompt inserting.~\gu{this picture does not seem to be important.}}
%     \label{fig:prompt_position}
% \end{figure}
The number of prompt tokens ($k$) is fixed to 10.
We train the model with 700 Java code snippets and evaluate the model on the code clone detection task of Solidity. 

\begin{figure}[tb]
    \centering
    \subfigure[\scriptsize{Ablation on the position of prompts}]{
        \includegraphics[scale = 0.15, trim=10 10 10 10]{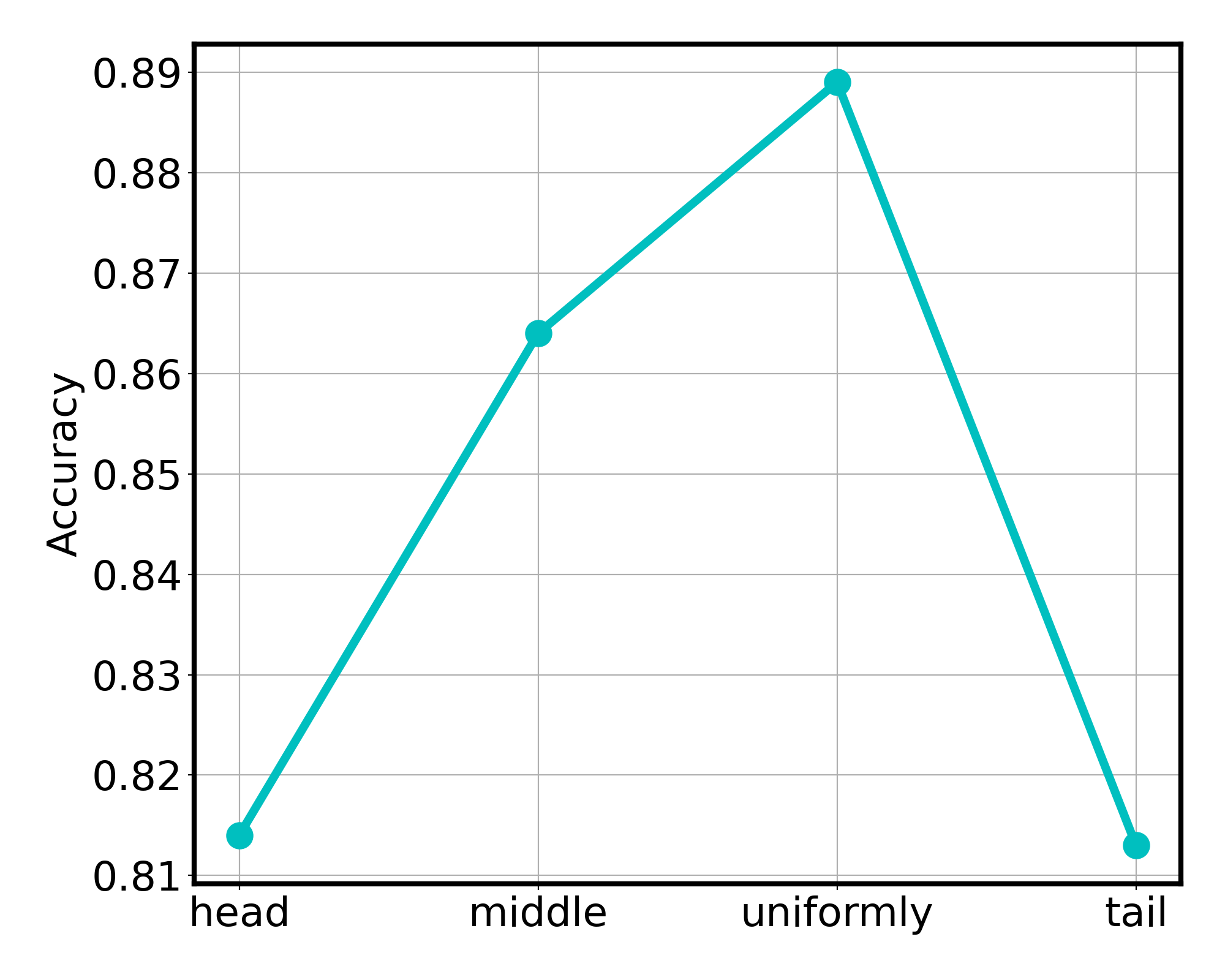}
        \label{fig:promptpos}
    }    
    \subfigure[\scriptsize{Ablation on the number of prompts}]{
        \includegraphics[scale = 0.15, trim=10 10 10 10]{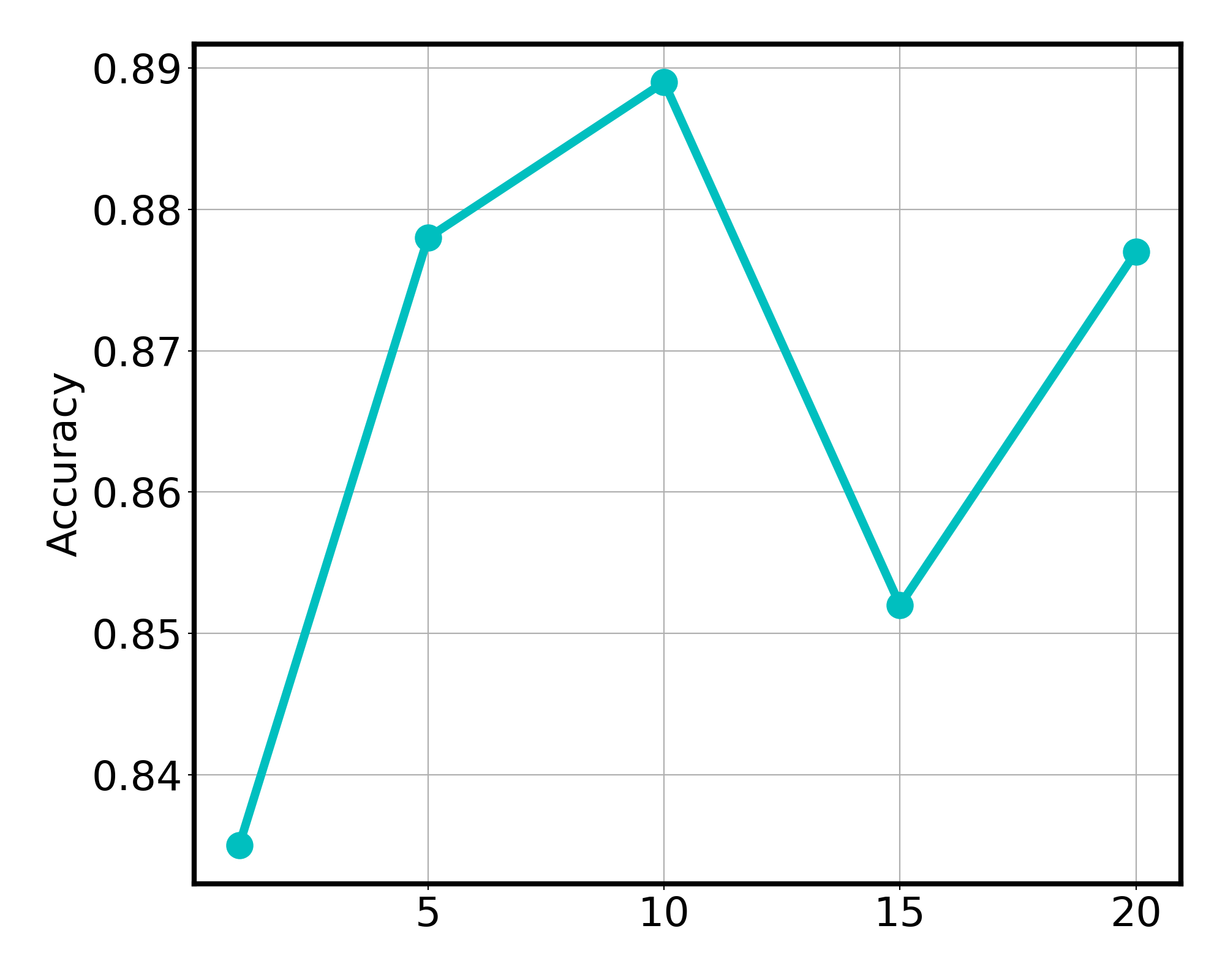}
        \label{fig:promptnum}
    }

    \caption{Performance of \approach under different prompt positions and prompt numbers on the Solidity clone detection dataset (SCCD).}
\end{figure}

As shown in Figure~\ref{fig:promptpos}, placing prompt tokens uniformly achieves the best performance compared to other templates. The reason could be that prompts have more influence to nearby tokens. By placing prompts uniformly, every input token can be influenced by sufficient prompt tokens.

\smallskip\noindent\textbf{Number of Prompt Tokens:}
We further assess the impact of prompt numbers. We insert prompt tokens uniformly into the PLM input and vary the number of prompt tokens~$k$ from 1 to 20. We train the model with 700 Java code snippets and evaluate it on the code clone detection task of Solidity.

As Figure~\ref{fig:promptnum} shows, the number of prompt tokens is strongly correlated to the performance of representation learning. Fewer prompt tokens can be insufficient to steer the PLM to yield meaningful prediction, while large numbers of prompts can restrict the input size. The optimal number of prompt tokens is 10 in our experiments.

\begin{figure}[tb]
\centerline{\includegraphics[width=0.5\textwidth]{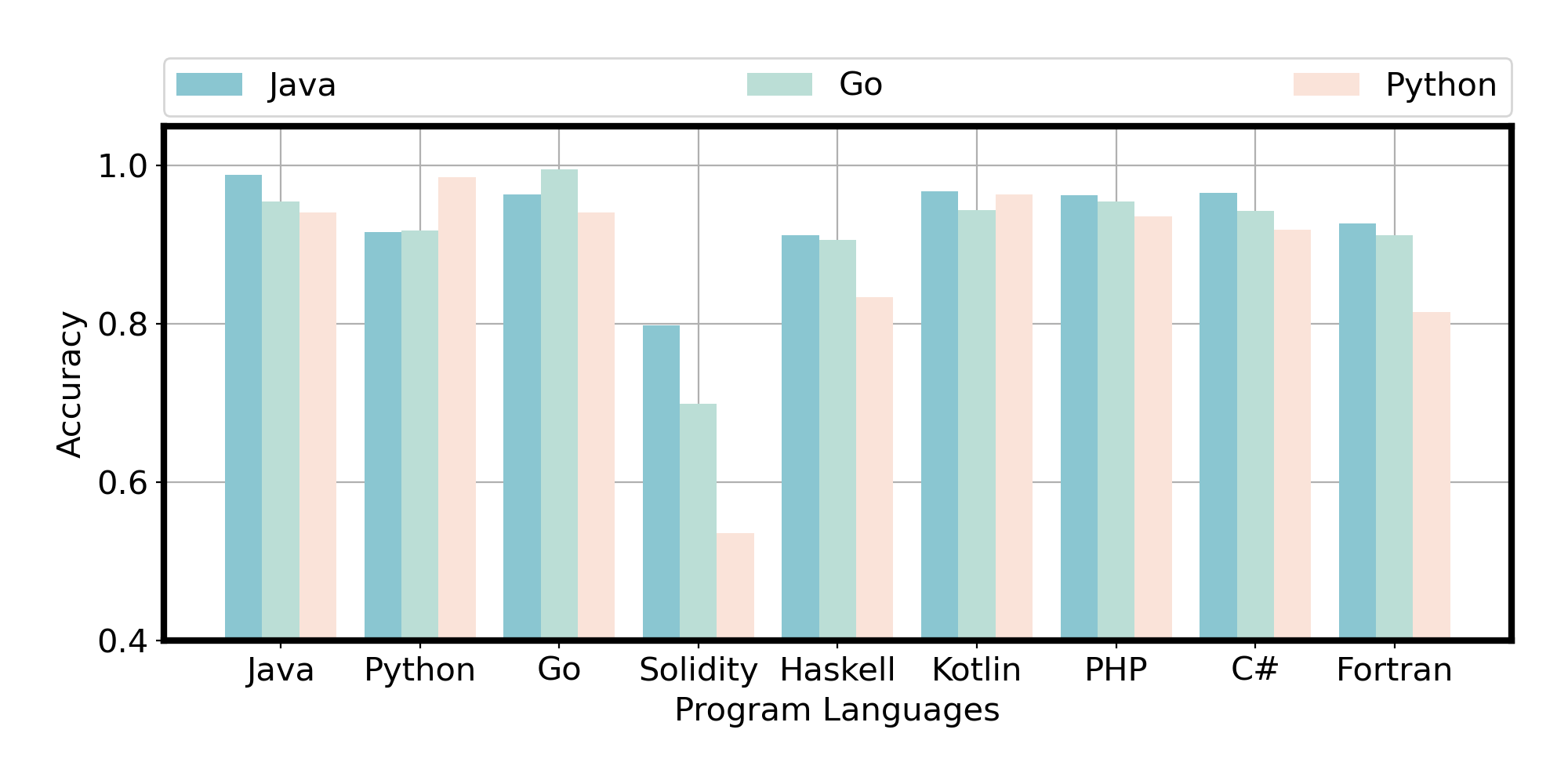}}
\caption{Performance of \approach in the code clone detection task with different source languages.}
\label{fig:diffsource}
\end{figure}

\smallskip\noindent\textbf{Source Languages:} 
To study the impact of different source languages, we train the model using 5,000 data samples of Java, Python, and C++, respectively. We evaluate the performance of zero-shot program representation on the code clone detection task of nine target languages in the CodeNet.
Figure~\ref{fig:diffsource} shows the results. We can observe that using Java as the source language achieves the best performance. This can be attributed to two reasons. First, CodeBERT is pre-trained using Java. Second, as a common language, Java contains more general features of programming languages compared to other languages. This facilitates the transfer of model to other languages with zero- or few-shot samples. 

\begin{figure*}[!t]
    \centerline{\includegraphics[width=0.99\textwidth]{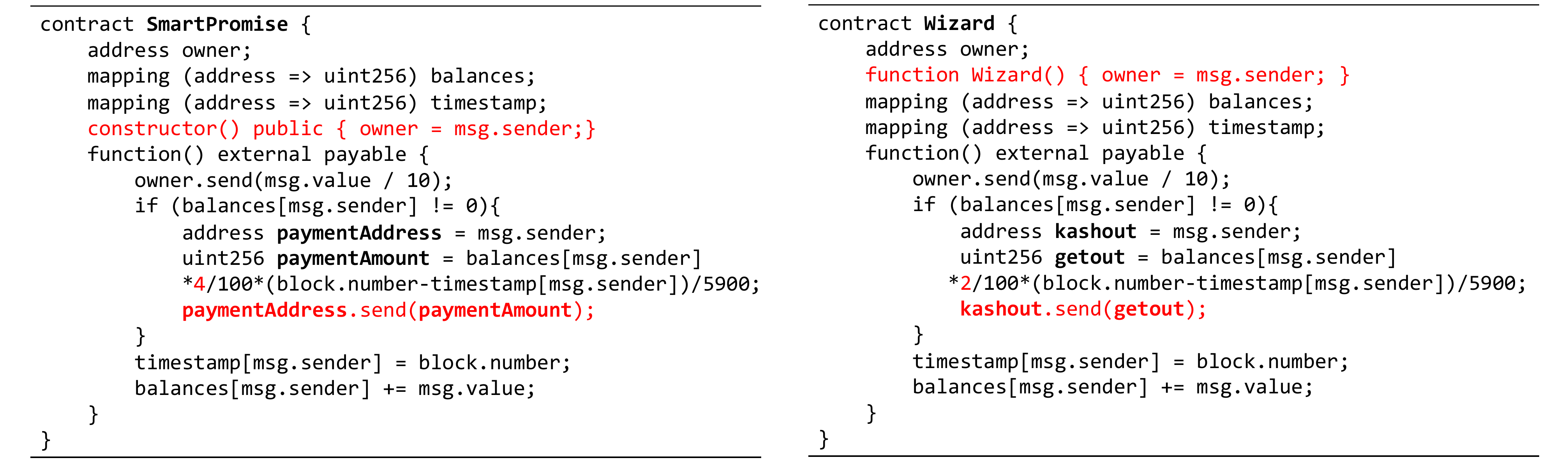}}
    \caption{An example of two cloned snippets in Solidity. \approach can successfully identify that the given snippets as cloned while CodeBERT cannot.}
    \label{fig:case_study}
\end{figure*}

\smallskip\noindent\textbf{PLM Scales:} 
Lastly, we study the effect of PLM scales. As Table~\ref{tab:rq1} and \ref{tab:rq3} indicate, larger PLM scale has a negative effect on the performance. For example, RoBERTa-large is almost twice the size of RoBERTa, but the latter performs even worse than the former. This is because large PLMs can easily overfit few data samples in the zero- or few-shot learning. 

%\begin{framed}\noindent
\emph{Answer to RQ4:}
The effectiveness of our approach is affected by prompt templates, source languages and PLM scales. Inserting ten prompt tokens uniformly to the original PLM input can steer the PLM to output better representations. Java as the source language can be better generalized to other languages. PLMs with a moderate scale can better fit for the zero- or few-shot learning.
%our hyperparameters selection is the optimal for the performance.
%\end{framed}	

%\section{Visualization of Program Representations}	
%	\gu{maybe for the next submission, we can visualize the representations (CLS vectors) and give some discussions.}

\subsection{Example}

\begin{table*}[!t]
\centering
\footnotesize
\caption{Models for Learning Program Representations.}
\label{tab:related1}
\begin{threeparttable}
\begin{tabular}{ccccc|cc|cccc}
\hline
\multirow{2}{*}{Model} & \multicolumn{4}{c|}{Training data}               & \multicolumn{2}{c|}{Generalization} & \multicolumn{4}{c}{Task adaption}           \\ \cline{2-11} 
                       & Unsupervised & Supervised & Few-shot & Zero-shot & General       & Task-specific      & None & Task-specific & Fine-tuning & Prompt \\ \hline
Code2vec \cite{code2vec}               & \checkmark            &            &          &           & \checkmark             &                    & \checkmark    &               &             &        \\
Code2seq \cite{code2seq}               & \checkmark            &            &          &           & \checkmark             &                    & \checkmark    &               &             &        \\
InferCode \cite{infercode}              & \checkmark            &            &          &           & \checkmark             &                    & \checkmark    &               &             &        \\ \hline
CD \cite{FangLS0S20}                     &              & \checkmark          &          &           &               & \checkmark                  &      & \checkmark             &             &        \\
DCRL \cite{ZhangHZWLS21}                   &              & \checkmark          &          &           &               & \checkmark                  &      & \checkmark             &             &        \\
CM \cite{ChoiBNL21}              &              & \checkmark          &          &           &               & \checkmark                  &      & \checkmark             &             &        \\
CS \cite{gu2018deepcs, HaldarWXH20, Liw20}                     &              & \checkmark          &          &           &               & \checkmark                  &      & \checkmark             &             &        \\
MNP \cite{ZhangCLP21}                    &              & \checkmark          &          &           &               & \checkmark                  &      & \checkmark             &             &        \\
DD \cite{ZhouLSD019}            &              & \checkmark          &          &           &               & \checkmark                  &      & \checkmark             &             &        \\
CC \cite{WangL21a}        &              & \checkmark          &          &           &               & \checkmark                  &      & \checkmark             &             &        \\ \hline
CodeBERT \cite{codebert}               &              &            & \checkmark        &           & \checkmark             &                    &      &               & \checkmark           &        \\
CodeT5 \cite{codet5}                 &              &            & \checkmark        &           & \checkmark             &                    &      &               & \checkmark           &        \\ 
\hline
\approach                &              &            &          & \checkmark         & \checkmark             &                    &      &               &             & \checkmark      \\ \hline
\end{tabular}
\begin{tablenotes}
\item * CD = clone detection, CM = code summarization, CS = code search, MNP = method name prediction, DD = defect detection, CC = code completion.
%\vspace{-4mm}
\end{tablenotes}
\end{threeparttable}
\end{table*}

We now provide a concrete example to demonstrate the effectiveness of \approach. Figure~\ref{fig:case_study} shows two cloned code snippets in Solidity. They are similar in functionality but different in key words and structures. For example, ``paymentAddress'' and ``kashout'' (highlighted in red) are two equivalent keywords in the two snippets. Because the two words are both domain specific, baseline models such as CodeBERT can hardly detect the clone without prior knowledge. Comparatively, \approach successfully detects the clone by reusing prior knowledge from PLMs using prompt learning. 
The prompts in \approach cast the underlying meaning in the PLM to downstream tasks, which help large PLMs capture word semantics even without training data.
%It can resolve the variables different in words but same in semantic and yield result that they are similar. Baselines like CodeBERT may only focus on the word similarity in the case of lacking training data and offer the wrong answer.

%\subsection{\cn{Implications}}

%\cn{\approach can be used in the same pipeline as traditional fine-tuning models except for the additional prompt tokens we add to the input during training and prediction. For programming languages which have few labeled samples, \approach is useful that it makes the most use of knowledge from the model trained by languages with sufficient training data through prompt-based learning. Languages like Solidity can perform well without labeled samples based on \approach in zero-shot or few-shot scenario.}

% \section{Discussions}
% \subsection{Limitations}
% We evaluate \approach on small data set which size is up to 5000. This size is quite small compared to normal way need such as fine-tuning. \approach performs well in this scenario while it can barely reach the same accuracy of fine-tuning when the data is sufficient. Many works \cite{prefixtuning, softprompt} related to prompt also aware of this and use ways such as limiting data set size or freezing PLM parameter for training to highlight the effectiveness of prompt in the scenario of few-shot learning or lightweight deployment.

% Although trainable prompts cost much less memory for saving parameter compared with fine-tuning header, it is also a little harder for searching the global optima in gradient descent, which costs about 30\% more time than fine-tuning. 

\section{Threats to Validity}

\textit{Internal Validity.}
Our approach is built upon CodeBERT. Although CodeBERT is the most popular PLM for learning program representations, other PLMs for code such as GPT-2 (unidirectional Transformers) and CodeT5 (with an encoder-decoder architecture) may have different results. However, we argue that our approach is independent on the PLM architecture itself since we merely modify the format of input and output of the PLM. %For example, for GPT-2 and CodeT5, we can simply take the first predicted token as the ``[MASK]'' token and map it to the final answer through a verbalizer. We leave the implements of zero-shot GPT-2 and CodeT5 for future work.

%\cn{Although languages like Java and Go are involved in the pre-training of CodeBERT and analyzed together with Solidity, we compare \approach with baselines in each programming language respectively. If our approach outperforms others in the specific language, it means our approach works.}

\smallskip\noindent\textit{External Validity.}
\approach is evaluated on classification tasks such as code clone detection. Other generative tasks such as code summarization might have different performance. However, prompt-based learning has also been shown to be effective in generative tasks~\cite{prefixtuning}. We leave the extensions of our approach to generative tasks for our future work.
In our work, the downstream tasks are assumed to be binary classifications. Hence, we represent the binary answers using two candidate words. We can use more candidate words for multi-class classification tasks. The candidate words are manually selected and can be searched to find the most suitable ones. 

%As for the selection of candidate words that verbalizer maps to the final answer of downstream tasks, in our approach we choose ``yes’‘ and ``no’‘. Actually it is not so import that what the specific word we choose. We can also choose ``relevant’‘ and ``irrelevant" as candidates too. What is really important is that the candidate words should be distinguished with each other to make MLM header give a better choice of which word is more likely for the ``[MASK]’‘ token. Also, the candidate words can also become trainable vectors just like prompts in our approach following \citet{warp}.

%In Smart Contract Clone Detection data set, because the Solidity data set for code clone detection is rare, we build the data set by ourselves and use some heuristic methods following \citet{bigclonebench} to save cost of human annotation. This may lead to the inaccuracy of the data set itself while the impact seems small as the experiment result showing the good performance.

\section{Related Work}
\label{sec:related}
\subsection{Learning Program Representations}

As the core prerequisite for many code intelligence tasks, learning program representations has been extensively explored in software engineering~\cite{codexglue}. Table~\ref{tab:related1} shows typical approaches in learning program representations. Broadly, they can be classified into three categories, %according to different criteria, 
including unsupervised for general languages, supervised for specific tasks, and few-shot learning.

%%% Unsupervised
The most typical category of work lie in the unsupervised approaches such as code2vec~\cite{code2vec}, code2seq~\cite{code2seq}, and InferCode~\cite{infercode}. Code2vec and code2seq aggregate representations of each path in AST (abstract syntax tree) based on attention. InferCode predicts subtrees automatically identified from the contexts of an AST in a self-supervised manner.
These methods directly learn program representation from AST paths. They utilize the word embedding techniques in natural language processing and incorporate them with semantic and syntax information in program source code.
%AST is a kind of special representation for program languages, which contains more standard semantic and syntax information in the format of a tree. Learning code vectors based on AST can make the most of unique characteristic in program languages.
The limitation of these methods is the lack of adaptions to downstream tasks. 
%For example, code2vec focuses on learning ASTs. Besides, 
The learned code vectors are fixed and cannot be fine-tuned on downstream tasks.
Furthermore, these methods are purely trained on code, thus are unsuitable for NL-PL tasks such as code search.

%%% supervised with downstream tasks
To improve the performance of downstream tasks, researchers have also resorted to task-oriented supervised learning methods~\cite{codenet}. 
For example, for code clone detection task, \citet{FangLS0S20} caught the similarity of semantics between two code snippets using a supervised deep learning model, which pays attention to caller-callee relationships and learns the hidden syntactic and semantic features of source codes. 
\citet{ZhangHZWLS21} disentangled the representation of semantic and syntax with AST and GAN (generative adversarial network), then used only semantic representation to detect code clone.
%Learning program representations is also a core component for code search and code summarization~\cite{ChoiBNL21}. Code search aims to capture the relationship between programming and natural languages. 
%For code search and code summarization tasks, \citet{HaldarWXH20} proposed a multi-perspective cross-lingual neural framework for code–text matching which can capture both global and local similarities between programming language and natural language. 
For code search task,
\citet{gu2018deepcs} proposed a code representation model named CODEnn to learn semantic representations of code snippets through jointly embedding with comments,
\citet{HaldarWXH20} designed a multi-perspective cross-lingual neural framework, and \citet{Liw20} learned code-query interactions. 
% \citet{ChoiBNL21} also introduces transformer and graph convolution network to effectively capture sequential and structural information for learning program representation.
\citet{ZhangCLP21} proposed a hybrid code representation learning approach to resolve program dependence and semantics for predicting method name.
\citet{YangCZSZ21} learned a unified vector representation of both methods and bug reports for method-level fault localization.
\citet{ZhouLSD019} constructed a graph neural network to learn semantic representation of code to identify the vulnerable functions.
\citet{WangL21a} proposed AST Graph Attention Block to capture different dependencies in the AST graph for representation learning in code completion.
These models are trained for the specific downstream tasks, which achieve good performance but lack generality to support multiple tasks with one single model. 

%%% PLMs and fine-tuning
The aforementioned methods require a large scale corpus to train the program representation model.
To alleviate this problem, pre-trained programming language models are proposed such as CodeBERT~\cite{codebert} and CodeT5~\cite{codet5}. It is a fine-tuning based few-shot program learning paradigm: PLMs learn a vast amount of knowledge from large scale unlabelled corpora in the pre-training phase, and achieve state-of-the-art accuracy in the fine-tuning phase with a small amount of labelled task-specific data.
This gives PLMs the basic generalization ability to handle a wide range of downstream tasks well. Task adaption through fine-tuning adds extra knowledge of specific tasks to PLMs and improves the performance.
However, in this paradigm, the gap between the pre-training phase and the downstream task can be significant: the objectives are different, and for the downstream tasks, we usually need to introduce new parameters.

%While all of these methods cost large scale training data. There are works such as \cite{HuangTSG0J0D20} that try to improve the model performance by annotating more data by human, but it is still too expensive to adapt for all downstream tasks without enough data set. 

%%% Our work
To the best of our knowledge, our \approach is the first zero-shot learning method for program representation. 
\approach follows a prompt-based learning paradigm for task adaption.
Prompt learning makes it possible for downstream tasks to take the same format as the pre-training objectives and require no new parameters. 
By narrowing the gap between the two phases, deploying the PLMs on specific tasks becomes much easier with little training data.

%\approach extracts the knowledge from PLMs more efficiently since it transforms the input into the format which is like that of MLM in pre-training phase. 
%Also, the ability of prompt and combined input can make PLMs learn better representation without any target program language corpora.
%\textit{Compared with these works, \approach requires little data set for training the model and makes use of PLMs in an efficient way through prompt-based learning. It can solve a wide range of code intelligence tasks instead of the specific task, showing good generalizability.}

%\subsection{Zero-shot Program Representations}
\subsection{Prompt-based Learning}
\label{sec:zeroshotlearning}
% In recent years, there are some zero-shot learning works in other fields, such as computer vision \cite{LiXWD21, WangCM0L21}. 

%In program representation learning, special features like AST in program can be used for zero-shot representation learning \cite{ChenA21} because they may contain more formatting information than natural language which can be read by machine more easily. But these methods may lack of generalization across program languages which use different AST rules. 

As a promising method for zero-shot learning, a growing number of prompt-based learning approaches~\cite{abs210713586} have been proposed in recent years. For example, \citet{pet} proposed PET which transforms the classification task into an MLM task and uses prompt to elicit knowledge from PLM. But the prompt is manually crafted and hard to select the most suitable words for it. \citet{autoprompt} proposed AutoPrompt which automatically searches prompt words discretely using gradient signals in the target task. Although discrete searching retains the semantic of prompt, it also cannot find out the most precise prompts for machine models. For solving this problem, \citet{prefixtuning} proposed Prefix-Tuning which optimizes a continuous task-specific vector prepended to every layer of the Transformer~\cite{transformer} in PLM and freezes the PLM for saving computation cost. The performance of Prefix-Tuning is excellent but it only focuses on natural language generation tasks. 

Comparatively, \approach optimizes the prompt vectors in continuous space instead of discrete words or human-writing, making the prompt more suitable for PLMs to understand and more efficient for extracting knowledge. Moreover, \approach is the first prompt method to solve programming language understanding tasks.
%, while others focus on the field of natural language or the generation tasks such as machine translation. 
%\approach is the most suitable one for program understanding tasks in zero-shot scenario among all of these prompt-based learning approaches.

\section{Conclusion}
\label{sec:conclusion}
In this paper, we propose \approach, a novel approach for  zero-shot program representation learning via prompt tuning. \approach improves traditional pre-trained programming language models by introducing prompt into program representation learning. %and optimizes the input structure. 
Experiments show that \approach outperforms baseline models in three code intelligence tasks, including code clone detection, code search and method name prediction in both zero-shot and few-shot settings. Program representations learned by \approach also demonstrate good generalizability for low/zero-resource programming languages.
In the future, we will investigate our approach in more languages and software engineering tasks.
%we will investigate more efficient methods for few-shot learning on PLMs.

Source code and datasets to reproduce our work are available at: https://github.com/ChrisCN97/zecoler.

%%
%% The acknowledgments section is defined using the "acks" environment
%% (and NOT an unnumbered section). This ensures the proper
%% identification of the section in the article metadata, and the
%% consistent spelling of the heading.
\begin{acks}
This research is supported by National Natural Science Foundation of China (Grant No. 62032004, 62102244).
\end{acks}

%%
%% The next two lines define the bibliography style to be used, and
%% the bibliography file.
\balance
%%% -*-BibTeX-*-
%%% Do NOT edit. File created by BibTeX with style
%%% ACM-Reference-Format-Journals [18-Jan-2012].

\bibliographystyle{ACM-Reference-Format}
\bibliography{references}

\begin{thebibliography}{44}

%%% ====================================================================
%%% NOTE TO THE USER: you can override these defaults by providing
%%% customized versions of any of these macros before the \bibliography
%%% command.  Each of them MUST provide its own final punctuation,
%%% except for \shownote{}, \showDOI{}, and \showURL{}.  The latter two
%%% do not use final punctuation, in order to avoid confusing it with
%%% the Web address.
%%%
%%% To suppress output of a particular field, define its macro to expand
%%% to an empty string, or better, \unskip, like this:
%%%
%%% \newcommand{\showDOI}[1]{\unskip}   % LaTeX syntax
%%%
%%% \def \showDOI #1{\unskip}           % plain TeX syntax
%%%
%%% ====================================================================

\ifx \showCODEN    \undefined \def \showCODEN     #1{\unskip}     \fi
\ifx \showDOI      \undefined \def \showDOI       #1{#1}\fi
\ifx \showISBNx    \undefined \def \showISBNx     #1{\unskip}     \fi
\ifx \showISBNxiii \undefined \def \showISBNxiii  #1{\unskip}     \fi
\ifx \showISSN     \undefined \def \showISSN      #1{\unskip}     \fi
\ifx \showLCCN     \undefined \def \showLCCN      #1{\unskip}     \fi
\ifx \shownote     \undefined \def \shownote      #1{#1}          \fi
\ifx \showarticletitle \undefined \def \showarticletitle #1{#1}   \fi
\ifx \showURL      \undefined \def \showURL       {\relax}        \fi
% The following commands are used for tagged output and should be
% invisible to TeX
\providecommand\bibfield[2]{#2}
\providecommand\bibinfo[2]{#2}
\providecommand\natexlab[1]{#1}
\providecommand\showeprint[2][]{arXiv:#2}

\bibitem[\protect\citeauthoryear{Ahmad, Chakraborty, Ray, and Chang}{Ahmad
  et~al\mbox{.}}{2021}]%
        {plbart}
\bibfield{author}{\bibinfo{person}{Wasi~Uddin Ahmad}, \bibinfo{person}{Saikat
  Chakraborty}, \bibinfo{person}{Baishakhi Ray}, {and}
  \bibinfo{person}{Kai{-}Wei Chang}.} \bibinfo{year}{2021}\natexlab{}.
\newblock \showarticletitle{Unified Pre-training for Program Understanding and
  Generation}. In \bibinfo{booktitle}{\emph{Proceedings of NAACL-HLT}}.
  \bibinfo{pages}{2655--2668}.
\newblock


\bibitem[\protect\citeauthoryear{Alon, Brody, Levy, and Yahav}{Alon
  et~al\mbox{.}}{2019a}]%
        {code2seq}
\bibfield{author}{\bibinfo{person}{Uri Alon}, \bibinfo{person}{Shaked Brody},
  \bibinfo{person}{Omer Levy}, {and} \bibinfo{person}{Eran Yahav}.}
  \bibinfo{year}{2019}\natexlab{a}.
\newblock \showarticletitle{code2seq: Generating Sequences from Structured
  Representations of Code}. In \bibinfo{booktitle}{\emph{Proceedings of ICLR}}.
\newblock


\bibitem[\protect\citeauthoryear{Alon, Zilberstein, Levy, and Yahav}{Alon
  et~al\mbox{.}}{2019b}]%
        {code2vec}
\bibfield{author}{\bibinfo{person}{Uri Alon}, \bibinfo{person}{Meital
  Zilberstein}, \bibinfo{person}{Omer Levy}, {and} \bibinfo{person}{Eran
  Yahav}.} \bibinfo{year}{2019}\natexlab{b}.
\newblock \showarticletitle{code2vec: learning distributed representations of
  code}.
\newblock \bibinfo{journal}{\emph{Proc. {ACM} Program. Lang.}}
  \bibinfo{volume}{3}, \bibinfo{number}{{POPL}} (\bibinfo{year}{2019}),
  \bibinfo{pages}{40:1--40:29}.
\newblock


\bibitem[\protect\citeauthoryear{Bengio, Courville, and Vincent}{Bengio
  et~al\mbox{.}}{2013}]%
        {BengioCV13}
\bibfield{author}{\bibinfo{person}{Yoshua Bengio}, \bibinfo{person}{Aaron~C.
  Courville}, {and} \bibinfo{person}{Pascal Vincent}.}
  \bibinfo{year}{2013}\natexlab{}.
\newblock \showarticletitle{Representation Learning: {A} Review and New
  Perspectives}.
\newblock \bibinfo{journal}{\emph{{IEEE} Trans. Pattern Anal. Mach. Intell.}}
  \bibinfo{volume}{35}, \bibinfo{number}{8} (\bibinfo{year}{2013}),
  \bibinfo{pages}{1798--1828}.
\newblock


\bibitem[\protect\citeauthoryear{Bornea, Pan, Rosenthal, Florian, and
  Sil}{Bornea et~al\mbox{.}}{2021}]%
        {BorneaPRFS21}
\bibfield{author}{\bibinfo{person}{Mihaela~A. Bornea}, \bibinfo{person}{Lin
  Pan}, \bibinfo{person}{Sara Rosenthal}, \bibinfo{person}{Radu Florian}, {and}
  \bibinfo{person}{Avirup Sil}.} \bibinfo{year}{2021}\natexlab{}.
\newblock \showarticletitle{Multilingual Transfer Learning for {QA} using
  Translation as Data Augmentation}. In \bibinfo{booktitle}{\emph{Proceedings
  of AAAI}}. \bibinfo{pages}{12583--12591}.
\newblock


\bibitem[\protect\citeauthoryear{Brown, Mann, Ryder, Subbiah, Kaplan, Dhariwal,
  Neelakantan, Shyam, Sastry, Askell, Agarwal, Herbert{-}Voss, Krueger,
  Henighan, Child, Ramesh, Ziegler, Wu, Winter, Hesse, Chen, Sigler, Litwin,
  Gray, Chess, Clark, Berner, McCandlish, Radford, Sutskever, and Amodei}{Brown
  et~al\mbox{.}}{2020}]%
        {gpt3}
\bibfield{author}{\bibinfo{person}{Tom~B. Brown}, \bibinfo{person}{Benjamin
  Mann}, \bibinfo{person}{Nick Ryder}, \bibinfo{person}{Melanie Subbiah},
  \bibinfo{person}{Jared Kaplan}, \bibinfo{person}{Prafulla Dhariwal},
  \bibinfo{person}{Arvind Neelakantan}, \bibinfo{person}{Pranav Shyam},
  \bibinfo{person}{Girish Sastry}, \bibinfo{person}{Amanda Askell},
  \bibinfo{person}{Sandhini Agarwal}, \bibinfo{person}{Ariel Herbert{-}Voss},
  \bibinfo{person}{Gretchen Krueger}, \bibinfo{person}{Tom Henighan},
  \bibinfo{person}{Rewon Child}, \bibinfo{person}{Aditya Ramesh},
  \bibinfo{person}{Daniel~M. Ziegler}, \bibinfo{person}{Jeffrey Wu},
  \bibinfo{person}{Clemens Winter}, \bibinfo{person}{Christopher Hesse},
  \bibinfo{person}{Mark Chen}, \bibinfo{person}{Eric Sigler},
  \bibinfo{person}{Mateusz Litwin}, \bibinfo{person}{Scott Gray},
  \bibinfo{person}{Benjamin Chess}, \bibinfo{person}{Jack Clark},
  \bibinfo{person}{Christopher Berner}, \bibinfo{person}{Sam McCandlish},
  \bibinfo{person}{Alec Radford}, \bibinfo{person}{Ilya Sutskever}, {and}
  \bibinfo{person}{Dario Amodei}.} \bibinfo{year}{2020}\natexlab{}.
\newblock \showarticletitle{Language Models are Few-Shot Learners}. In
  \bibinfo{booktitle}{\emph{Proceedings of NeurIPS}}.
  \bibinfo{pages}{1877--1901}.
\newblock


\bibitem[\protect\citeauthoryear{Bui, Yu, and Jiang}{Bui et~al\mbox{.}}{2021}]%
        {infercode}
\bibfield{author}{\bibinfo{person}{Nghi D.~Q. Bui}, \bibinfo{person}{Yijun Yu},
  {and} \bibinfo{person}{Lingxiao Jiang}.} \bibinfo{year}{2021}\natexlab{}.
\newblock \showarticletitle{InferCode: Self-Supervised Learning of Code
  Representations by Predicting Subtrees}. In
  \bibinfo{booktitle}{\emph{Proceedings of ICSE}}. \bibinfo{pages}{1186--1197}.
\newblock


\bibitem[\protect\citeauthoryear{Chen, Liao, Zhang, Huang, and Zheng}{Chen
  et~al\mbox{.}}{2021}]%
        {ChenLZ0Z21}
\bibfield{author}{\bibinfo{person}{Xiangping Chen}, \bibinfo{person}{Peiyong
  Liao}, \bibinfo{person}{Yixin Zhang}, \bibinfo{person}{Yuan Huang}, {and}
  \bibinfo{person}{Zibin Zheng}.} \bibinfo{year}{2021}\natexlab{}.
\newblock \showarticletitle{Understanding Code Reuse in Smart Contracts}. In
  \bibinfo{booktitle}{\emph{Proceedings of ({SANER})}}.
  \bibinfo{pages}{470--479}.
\newblock


\bibitem[\protect\citeauthoryear{Choi, Bak, Na, and Lee}{Choi
  et~al\mbox{.}}{2021}]%
        {ChoiBNL21}
\bibfield{author}{\bibinfo{person}{YunSeok Choi}, \bibinfo{person}{JinYeong
  Bak}, \bibinfo{person}{CheolWon Na}, {and} \bibinfo{person}{Jee{-}Hyong
  Lee}.} \bibinfo{year}{2021}\natexlab{}.
\newblock \showarticletitle{Learning Sequential and Structural Information for
  Source Code Summarization}. In \bibinfo{booktitle}{\emph{Proceedings of
  ACL/IJCNLP}}. \bibinfo{pages}{2842--2851}.
\newblock


\bibitem[\protect\citeauthoryear{Compton, Frank, Patros, and Koay}{Compton
  et~al\mbox{.}}{2020}]%
        {ComptonFPK20}
\bibfield{author}{\bibinfo{person}{Rhys Compton}, \bibinfo{person}{Eibe Frank},
  \bibinfo{person}{Panos Patros}, {and} \bibinfo{person}{Abigail Koay}.}
  \bibinfo{year}{2020}\natexlab{}.
\newblock \showarticletitle{Embedding Java Classes with code2vec: Improvements
  from Variable Obfuscation}. In \bibinfo{booktitle}{\emph{Proceedings of
  MSR}}. \bibinfo{pages}{243--253}.
\newblock


\bibitem[\protect\citeauthoryear{Devlin, Chang, Lee, and Toutanova}{Devlin
  et~al\mbox{.}}{2019}]%
        {DevlinCLT19}
\bibfield{author}{\bibinfo{person}{Jacob Devlin}, \bibinfo{person}{Ming{-}Wei
  Chang}, \bibinfo{person}{Kenton Lee}, {and} \bibinfo{person}{Kristina
  Toutanova}.} \bibinfo{year}{2019}\natexlab{}.
\newblock \showarticletitle{{BERT:} Pre-training of Deep Bidirectional
  Transformers for Language Understanding}. In
  \bibinfo{booktitle}{\emph{Proceedings of NAACL-HLT}}.
  \bibinfo{pages}{4171--4186}.
\newblock


\bibitem[\protect\citeauthoryear{Fang, Liu, Shi, Huang, and Shi}{Fang
  et~al\mbox{.}}{2020}]%
        {FangLS0S20}
\bibfield{author}{\bibinfo{person}{Chunrong Fang}, \bibinfo{person}{Zixi Liu},
  \bibinfo{person}{Yangyang Shi}, \bibinfo{person}{Jeff Huang}, {and}
  \bibinfo{person}{Qingkai Shi}.} \bibinfo{year}{2020}\natexlab{}.
\newblock \showarticletitle{Functional code clone detection with syntax and
  semantics fusion learning}. In \bibinfo{booktitle}{\emph{Proceedings of
  ISSTA}}. \bibinfo{pages}{516--527}.
\newblock


\bibitem[\protect\citeauthoryear{Feng, Guo, Tang, Duan, Feng, Gong, Shou, Qin,
  Liu, Jiang, and Zhou}{Feng et~al\mbox{.}}{2020}]%
        {codebert}
\bibfield{author}{\bibinfo{person}{Zhangyin Feng}, \bibinfo{person}{Daya Guo},
  \bibinfo{person}{Duyu Tang}, \bibinfo{person}{Nan Duan},
  \bibinfo{person}{Xiaocheng Feng}, \bibinfo{person}{Ming Gong},
  \bibinfo{person}{Linjun Shou}, \bibinfo{person}{Bing Qin},
  \bibinfo{person}{Ting Liu}, \bibinfo{person}{Daxin Jiang}, {and}
  \bibinfo{person}{Ming Zhou}.} \bibinfo{year}{2020}\natexlab{}.
\newblock \showarticletitle{CodeBERT: {A} Pre-Trained Model for Programming and
  Natural Languages}. In \bibinfo{booktitle}{\emph{Proceedings of EMNLP}}.
  \bibinfo{pages}{1536--1547}.
\newblock


\bibitem[\protect\citeauthoryear{Gu, Wang, Chen, Li, and Cho}{Gu
  et~al\mbox{.}}{2018a}]%
        {GuWCLC18}
\bibfield{author}{\bibinfo{person}{Jiatao Gu}, \bibinfo{person}{Yong Wang},
  \bibinfo{person}{Yun Chen}, \bibinfo{person}{Victor O.~K. Li}, {and}
  \bibinfo{person}{Kyunghyun Cho}.} \bibinfo{year}{2018}\natexlab{a}.
\newblock \showarticletitle{Meta-Learning for Low-Resource Neural Machine
  Translation}. In \bibinfo{booktitle}{\emph{Proceedings of EMNLP}}.
  \bibinfo{pages}{3622--3631}.
\newblock


\bibitem[\protect\citeauthoryear{Gu, Zhang, and Kim}{Gu et~al\mbox{.}}{2018b}]%
        {gu2018deepcs}
\bibfield{author}{\bibinfo{person}{Xiaodong Gu}, \bibinfo{person}{Hongyu
  Zhang}, {and} \bibinfo{person}{Sunghun Kim}.}
  \bibinfo{year}{2018}\natexlab{b}.
\newblock \showarticletitle{Deep code search}. In
  \bibinfo{booktitle}{\emph{Proceedings of ICSE}}. \bibinfo{pages}{933--944}.
\newblock


\bibitem[\protect\citeauthoryear{Haldar, Wu, Xiong, and Hockenmaier}{Haldar
  et~al\mbox{.}}{2020}]%
        {HaldarWXH20}
\bibfield{author}{\bibinfo{person}{Rajarshi Haldar}, \bibinfo{person}{Lingfei
  Wu}, \bibinfo{person}{Jinjun Xiong}, {and} \bibinfo{person}{Julia
  Hockenmaier}.} \bibinfo{year}{2020}\natexlab{}.
\newblock \showarticletitle{A Multi-Perspective Architecture for Semantic Code
  Search}. In \bibinfo{booktitle}{\emph{Proceedings of ACL}}.
  \bibinfo{pages}{8563--8568}.
\newblock


\bibitem[\protect\citeauthoryear{Husain, Wu, Gazit, Allamanis, and
  Brockschmidt}{Husain et~al\mbox{.}}{2019}]%
        {codesearchnet}
\bibfield{author}{\bibinfo{person}{Hamel Husain}, \bibinfo{person}{Ho{-}Hsiang
  Wu}, \bibinfo{person}{Tiferet Gazit}, \bibinfo{person}{Miltiadis Allamanis},
  {and} \bibinfo{person}{Marc Brockschmidt}.} \bibinfo{year}{2019}\natexlab{}.
\newblock \showarticletitle{CodeSearchNet Challenge: Evaluating the State of
  Semantic Code Search}.
\newblock \bibinfo{journal}{\emph{CoRR}}  \bibinfo{volume}{abs/1909.09436}
  (\bibinfo{year}{2019}).
\newblock


\bibitem[\protect\citeauthoryear{Li, Qin, Yan, Shen, and Chen}{Li
  et~al\mbox{.}}{2020}]%
        {Liw20}
\bibfield{author}{\bibinfo{person}{Wei Li}, \bibinfo{person}{Haozhe Qin},
  \bibinfo{person}{Shuhan Yan}, \bibinfo{person}{Beijun Shen}, {and}
  \bibinfo{person}{Yuting Chen}.} \bibinfo{year}{2020}\natexlab{}.
\newblock \showarticletitle{Learning Code-Query Interaction for Enhancing Code
  Searches}. In \bibinfo{booktitle}{\emph{Proceedings of ICSME}}.
  \bibinfo{publisher}{{IEEE}}, \bibinfo{pages}{115--126}.
\newblock


\bibitem[\protect\citeauthoryear{Li and Liang}{Li and Liang}{2021}]%
        {prefixtuning}
\bibfield{author}{\bibinfo{person}{Xiang~Lisa Li} {and} \bibinfo{person}{Percy
  Liang}.} \bibinfo{year}{2021}\natexlab{}.
\newblock \showarticletitle{Prefix-Tuning: Optimizing Continuous Prompts for
  Generation}. In \bibinfo{booktitle}{\emph{Proceedings of ACL/IJCNLP}}.
  \bibinfo{pages}{4582--4597}.
\newblock


\bibitem[\protect\citeauthoryear{Liu, Yuan, Fu, Jiang, Hayashi, and Neubig}{Liu
  et~al\mbox{.}}{2021a}]%
        {abs210713586}
\bibfield{author}{\bibinfo{person}{Pengfei Liu}, \bibinfo{person}{Weizhe Yuan},
  \bibinfo{person}{Jinlan Fu}, \bibinfo{person}{Zhengbao Jiang},
  \bibinfo{person}{Hiroaki Hayashi}, {and} \bibinfo{person}{Graham Neubig}.}
  \bibinfo{year}{2021}\natexlab{a}.
\newblock \showarticletitle{Pre-train, Prompt, and Predict: {A} Systematic
  Survey of Prompting Methods in Natural Language Processing}.
\newblock \bibinfo{journal}{\emph{CoRR}}  \bibinfo{volume}{abs/2107.13586}
  (\bibinfo{year}{2021}).
\newblock


\bibitem[\protect\citeauthoryear{Liu, Zheng, Du, Ding, Qian, Yang, and
  Tang}{Liu et~al\mbox{.}}{2021b}]%
        {ptuning}
\bibfield{author}{\bibinfo{person}{Xiao Liu}, \bibinfo{person}{Yanan Zheng},
  \bibinfo{person}{Zhengxiao Du}, \bibinfo{person}{Ming Ding},
  \bibinfo{person}{Yujie Qian}, \bibinfo{person}{Zhilin Yang}, {and}
  \bibinfo{person}{Jie Tang}.} \bibinfo{year}{2021}\natexlab{b}.
\newblock \showarticletitle{{GPT} Understands, Too}.
\newblock \bibinfo{journal}{\emph{CoRR}}  \bibinfo{volume}{abs/2103.10385}
  (\bibinfo{year}{2021}).
\newblock


\bibitem[\protect\citeauthoryear{Liu, Ott, Goyal, Du, Joshi, Chen, Levy, Lewis,
  Zettlemoyer, and Stoyanov}{Liu et~al\mbox{.}}{2019}]%
        {roberta}
\bibfield{author}{\bibinfo{person}{Yinhan Liu}, \bibinfo{person}{Myle Ott},
  \bibinfo{person}{Naman Goyal}, \bibinfo{person}{Jingfei Du},
  \bibinfo{person}{Mandar Joshi}, \bibinfo{person}{Danqi Chen},
  \bibinfo{person}{Omer Levy}, \bibinfo{person}{Mike Lewis},
  \bibinfo{person}{Luke Zettlemoyer}, {and} \bibinfo{person}{Veselin
  Stoyanov}.} \bibinfo{year}{2019}\natexlab{}.
\newblock \showarticletitle{RoBERTa: {A} Robustly Optimized {BERT} Pretraining
  Approach}.
\newblock \bibinfo{journal}{\emph{CoRR}}  \bibinfo{volume}{abs/1907.11692}
  (\bibinfo{year}{2019}).
\newblock


\bibitem[\protect\citeauthoryear{Loshchilov and Hutter}{Loshchilov and
  Hutter}{2017}]%
        {adamw}
\bibfield{author}{\bibinfo{person}{Ilya Loshchilov} {and}
  \bibinfo{person}{Frank Hutter}.} \bibinfo{year}{2017}\natexlab{}.
\newblock \showarticletitle{Fixing Weight Decay Regularization in Adam}.
\newblock \bibinfo{journal}{\emph{CoRR}}  \bibinfo{volume}{abs/1711.05101}
  (\bibinfo{year}{2017}).
\newblock


\bibitem[\protect\citeauthoryear{Lu, Guo, Ren, Huang, Svyatkovskiy, Blanco,
  Clement, Drain, Jiang, Tang, Li, Zhou, Shou, Zhou, Tufano, Gong, Zhou, Duan,
  Sundaresan, Deng, Fu, and Liu}{Lu et~al\mbox{.}}{2021}]%
        {codexglue}
\bibfield{author}{\bibinfo{person}{Shuai Lu}, \bibinfo{person}{Daya Guo},
  \bibinfo{person}{Shuo Ren}, \bibinfo{person}{Junjie Huang},
  \bibinfo{person}{Alexey Svyatkovskiy}, \bibinfo{person}{Ambrosio Blanco},
  \bibinfo{person}{Colin~B. Clement}, \bibinfo{person}{Dawn Drain},
  \bibinfo{person}{Daxin Jiang}, \bibinfo{person}{Duyu Tang},
  \bibinfo{person}{Ge Li}, \bibinfo{person}{Lidong Zhou},
  \bibinfo{person}{Linjun Shou}, \bibinfo{person}{Long Zhou},
  \bibinfo{person}{Michele Tufano}, \bibinfo{person}{Ming Gong},
  \bibinfo{person}{Ming Zhou}, \bibinfo{person}{Nan Duan},
  \bibinfo{person}{Neel Sundaresan}, \bibinfo{person}{Shao~Kun Deng},
  \bibinfo{person}{Shengyu Fu}, {and} \bibinfo{person}{Shujie Liu}.}
  \bibinfo{year}{2021}\natexlab{}.
\newblock \showarticletitle{CodeXGLUE: {A} Machine Learning Benchmark Dataset
  for Code Understanding and Generation}.
\newblock \bibinfo{journal}{\emph{CoRR}}  \bibinfo{volume}{abs/2102.04664}
  (\bibinfo{year}{2021}).
\newblock


\bibitem[\protect\citeauthoryear{Mahto, Vo, Turek, and Huth}{Mahto
  et~al\mbox{.}}{2021}]%
        {MahtoVTH21}
\bibfield{author}{\bibinfo{person}{Shivangi Mahto}, \bibinfo{person}{Vy~Ai Vo},
  \bibinfo{person}{Javier~S. Turek}, {and} \bibinfo{person}{Alexander Huth}.}
  \bibinfo{year}{2021}\natexlab{}.
\newblock \showarticletitle{Multi-timescale Representation Learning in {LSTM}
  Language Models}. In \bibinfo{booktitle}{\emph{Proceedings of ICLR}}.
\newblock


\bibitem[\protect\citeauthoryear{Puri, Kung, Janssen, Zhang, Domeniconi,
  Zolotov, Dolby, Chen, Choudhury, Decker, Thost, Buratti, Pujar, and
  Finkler}{Puri et~al\mbox{.}}{2021}]%
        {codenet}
\bibfield{author}{\bibinfo{person}{Ruchir Puri}, \bibinfo{person}{David~S.
  Kung}, \bibinfo{person}{Geert Janssen}, \bibinfo{person}{Wei Zhang},
  \bibinfo{person}{Giacomo Domeniconi}, \bibinfo{person}{Vladimir Zolotov},
  \bibinfo{person}{Julian Dolby}, \bibinfo{person}{Jie Chen},
  \bibinfo{person}{Mihir~R. Choudhury}, \bibinfo{person}{Lindsey Decker},
  \bibinfo{person}{Veronika Thost}, \bibinfo{person}{Luca Buratti},
  \bibinfo{person}{Saurabh Pujar}, {and} \bibinfo{person}{Ulrich Finkler}.}
  \bibinfo{year}{2021}\natexlab{}.
\newblock \showarticletitle{Project CodeNet: {A} Large-Scale {AI} for Code
  Dataset for Learning a Diversity of Coding Tasks}.
\newblock \bibinfo{journal}{\emph{CoRR}}  \bibinfo{volume}{abs/2105.12655}
  (\bibinfo{year}{2021}).
\newblock


\bibitem[\protect\citeauthoryear{Radford, Narasimhan, Salimans, and
  Sutskever}{Radford et~al\mbox{.}}{2018}]%
        {radford2018improving}
\bibfield{author}{\bibinfo{person}{Alec Radford}, \bibinfo{person}{Karthik
  Narasimhan}, \bibinfo{person}{Tim Salimans}, {and} \bibinfo{person}{Ilya
  Sutskever}.} \bibinfo{year}{2018}\natexlab{}.
\newblock \showarticletitle{Improving language understanding by generative
  pre-training}.
\newblock  (\bibinfo{year}{2018}).
\newblock


\bibitem[\protect\citeauthoryear{Raffel, Shazeer, Roberts, Lee, Narang, Matena,
  Zhou, Li, and Liu}{Raffel et~al\mbox{.}}{2020}]%
        {RaffelSRLNMZLL20}
\bibfield{author}{\bibinfo{person}{Colin Raffel}, \bibinfo{person}{Noam
  Shazeer}, \bibinfo{person}{Adam Roberts}, \bibinfo{person}{Katherine Lee},
  \bibinfo{person}{Sharan Narang}, \bibinfo{person}{Michael Matena},
  \bibinfo{person}{Yanqi Zhou}, \bibinfo{person}{Wei Li}, {and}
  \bibinfo{person}{Peter~J. Liu}.} \bibinfo{year}{2020}\natexlab{}.
\newblock \showarticletitle{Exploring the Limits of Transfer Learning with a
  Unified Text-to-Text Transformer}.
\newblock \bibinfo{journal}{\emph{J. Mach. Learn. Res.}}  \bibinfo{volume}{21}
  (\bibinfo{year}{2020}), \bibinfo{pages}{140:1--140:67}.
\newblock


\bibitem[\protect\citeauthoryear{Ravi and Larochelle}{Ravi and
  Larochelle}{2017}]%
        {metalearning1}
\bibfield{author}{\bibinfo{person}{Sachin Ravi} {and} \bibinfo{person}{Hugo
  Larochelle}.} \bibinfo{year}{2017}\natexlab{}.
\newblock \showarticletitle{Optimization as a Model for Few-Shot Learning}. In
  \bibinfo{booktitle}{\emph{Proceedings of ICLR}}.
\newblock


\bibitem[\protect\citeauthoryear{Salza, Schwizer, Gu, and Gall}{Salza
  et~al\mbox{.}}{2021}]%
        {salza2021effectiveness}
\bibfield{author}{\bibinfo{person}{Pasquale Salza}, \bibinfo{person}{Christoph
  Schwizer}, \bibinfo{person}{Jian Gu}, {and} \bibinfo{person}{Harald~C Gall}.}
  \bibinfo{year}{2021}\natexlab{}.
\newblock \showarticletitle{On the Effectiveness of Transfer Learning for Code
  Search}.
\newblock \bibinfo{journal}{\emph{arXiv preprint arXiv:2108.05890}}
  (\bibinfo{year}{2021}).
\newblock


\bibitem[\protect\citeauthoryear{Schick and Sch{\"{u}}tze}{Schick and
  Sch{\"{u}}tze}{2021}]%
        {pet}
\bibfield{author}{\bibinfo{person}{Timo Schick} {and} \bibinfo{person}{Hinrich
  Sch{\"{u}}tze}.} \bibinfo{year}{2021}\natexlab{}.
\newblock \showarticletitle{Exploiting Cloze-Questions for Few-Shot Text
  Classification and Natural Language Inference}. In
  \bibinfo{booktitle}{\emph{Proceedings of EACL}}. \bibinfo{pages}{255--269}.
\newblock


\bibitem[\protect\citeauthoryear{Shin, Razeghi, IV, Wallace, and Singh}{Shin
  et~al\mbox{.}}{2020}]%
        {autoprompt}
\bibfield{author}{\bibinfo{person}{Taylor Shin}, \bibinfo{person}{Yasaman
  Razeghi}, \bibinfo{person}{Robert L.~Logan IV}, \bibinfo{person}{Eric
  Wallace}, {and} \bibinfo{person}{Sameer Singh}.}
  \bibinfo{year}{2020}\natexlab{}.
\newblock \showarticletitle{AutoPrompt: Eliciting Knowledge from Language
  Models with Automatically Generated Prompts}. In
  \bibinfo{booktitle}{\emph{Proceedings of EMNLP}}.
  \bibinfo{pages}{4222--4235}.
\newblock


\bibitem[\protect\citeauthoryear{Snell, Swersky, and Zemel}{Snell
  et~al\mbox{.}}{2017}]%
        {metalearning2}
\bibfield{author}{\bibinfo{person}{Jake Snell}, \bibinfo{person}{Kevin
  Swersky}, {and} \bibinfo{person}{Richard~S. Zemel}.}
  \bibinfo{year}{2017}\natexlab{}.
\newblock \showarticletitle{Prototypical Networks for Few-shot Learning}. In
  \bibinfo{booktitle}{\emph{Proceedings of NeurIPS}}.
  \bibinfo{pages}{4077--4087}.
\newblock


\bibitem[\protect\citeauthoryear{Sun, Zheng, Hao, and Qiu}{Sun
  et~al\mbox{.}}{2021}]%
        {nspbert}
\bibfield{author}{\bibinfo{person}{Yi Sun}, \bibinfo{person}{Yu Zheng},
  \bibinfo{person}{Chao Hao}, {and} \bibinfo{person}{Hangping Qiu}.}
  \bibinfo{year}{2021}\natexlab{}.
\newblock \showarticletitle{{NSP-BERT:} {A} Prompt-based Zero-Shot Learner
  Through an Original Pre-training Task-Next Sentence Prediction}.
\newblock \bibinfo{journal}{\emph{CoRR}}  \bibinfo{volume}{abs/2109.03564}
  (\bibinfo{year}{2021}).
\newblock


\bibitem[\protect\citeauthoryear{Svajlenko, Islam, Keivanloo, Roy, and
  Mia}{Svajlenko et~al\mbox{.}}{2014}]%
        {bigclonebench}
\bibfield{author}{\bibinfo{person}{Jeffrey Svajlenko},
  \bibinfo{person}{Judith~F. Islam}, \bibinfo{person}{Iman Keivanloo},
  \bibinfo{person}{Chanchal~Kumar Roy}, {and} \bibinfo{person}{Mohammad~Mamun
  Mia}.} \bibinfo{year}{2014}\natexlab{}.
\newblock \showarticletitle{Towards a Big Data Curated Benchmark of
  Inter-project Code Clones}. In \bibinfo{booktitle}{\emph{Proceedings of
  ICSME}}. \bibinfo{pages}{476--480}.
\newblock


\bibitem[\protect\citeauthoryear{Vaswani, Shazeer, Parmar, Uszkoreit, Jones,
  Gomez, Kaiser, and Polosukhin}{Vaswani et~al\mbox{.}}{2017}]%
        {transformer}
\bibfield{author}{\bibinfo{person}{Ashish Vaswani}, \bibinfo{person}{Noam
  Shazeer}, \bibinfo{person}{Niki Parmar}, \bibinfo{person}{Jakob Uszkoreit},
  \bibinfo{person}{Llion Jones}, \bibinfo{person}{Aidan~N. Gomez},
  \bibinfo{person}{Lukasz Kaiser}, {and} \bibinfo{person}{Illia Polosukhin}.}
  \bibinfo{year}{2017}\natexlab{}.
\newblock \showarticletitle{Attention is All you Need}. In
  \bibinfo{booktitle}{\emph{Proceedings of NIPS}}. \bibinfo{pages}{5998--6008}.
\newblock


\bibitem[\protect\citeauthoryear{Wang and Li}{Wang and Li}{2021}]%
        {WangL21a}
\bibfield{author}{\bibinfo{person}{Yanlin Wang} {and} \bibinfo{person}{Hui
  Li}.} \bibinfo{year}{2021}\natexlab{}.
\newblock \showarticletitle{Code Completion by Modeling Flattened Abstract
  Syntax Trees as Graphs}. In \bibinfo{booktitle}{\emph{Proceedings of AAAI}}.
  \bibinfo{pages}{14015--14023}.
\newblock


\bibitem[\protect\citeauthoryear{Wang, Wang, Joty, and Hoi}{Wang
  et~al\mbox{.}}{2021}]%
        {codet5}
\bibfield{author}{\bibinfo{person}{Yue Wang}, \bibinfo{person}{Weishi Wang},
  \bibinfo{person}{Shafiq~R. Joty}, {and} \bibinfo{person}{Steven C.~H. Hoi}.}
  \bibinfo{year}{2021}\natexlab{}.
\newblock \showarticletitle{CodeT5: Identifier-aware Unified Pre-trained
  Encoder-Decoder Models for Code Understanding and Generation}. In
  \bibinfo{booktitle}{\emph{Proceedings of EMNLP}}.
  \bibinfo{pages}{8696--8708}.
\newblock


\bibitem[\protect\citeauthoryear{Wei and Li}{Wei and Li}{2017}]%
        {clonedetection}
\bibfield{author}{\bibinfo{person}{Huihui Wei} {and} \bibinfo{person}{Ming
  Li}.} \bibinfo{year}{2017}\natexlab{}.
\newblock \showarticletitle{Supervised Deep Features for Software Functional
  Clone Detection by Exploiting Lexical and Syntactical Information in Source
  Code}. In \bibinfo{booktitle}{\emph{Proceedings of IJCAI}}.
  \bibinfo{pages}{3034--3040}.
\newblock


\bibitem[\protect\citeauthoryear{Yang, Cao, Zeng, Shen, and Zhong}{Yang
  et~al\mbox{.}}{2021a}]%
        {YangCZSZ21}
\bibfield{author}{\bibinfo{person}{Shouliang Yang}, \bibinfo{person}{Junming
  Cao}, \bibinfo{person}{Hushuang Zeng}, \bibinfo{person}{Beijun Shen}, {and}
  \bibinfo{person}{Hao Zhong}.} \bibinfo{year}{2021}\natexlab{a}.
\newblock \showarticletitle{Locating Faulty Methods with a Mixed {RNN} and
  Attention Model}. In \bibinfo{booktitle}{\emph{Proceedings of ICPC}}.
  \bibinfo{pages}{207--218}.
\newblock


\bibitem[\protect\citeauthoryear{Yang, Keung, Yu, Gu, Wei, Ma, and Zhang}{Yang
  et~al\mbox{.}}{2021b}]%
        {YangKYGWMZ21}
\bibfield{author}{\bibinfo{person}{Zhen Yang}, \bibinfo{person}{Jacky Keung},
  \bibinfo{person}{Xiao Yu}, \bibinfo{person}{Xiaodong Gu},
  \bibinfo{person}{Zhengyuan Wei}, \bibinfo{person}{Xiaoxue Ma}, {and}
  \bibinfo{person}{Miao Zhang}.} \bibinfo{year}{2021}\natexlab{b}.
\newblock \showarticletitle{A Multi-Modal Transformer-based Code Summarization
  Approach for Smart Contracts}. In \bibinfo{booktitle}{\emph{Proceedings of
  ICPC}}. \bibinfo{pages}{1--12}.
\newblock


\bibitem[\protect\citeauthoryear{Zhang, Chen, Li, and Peng}{Zhang
  et~al\mbox{.}}{2021a}]%
        {ZhangCLP21}
\bibfield{author}{\bibinfo{person}{Fengyi Zhang}, \bibinfo{person}{Bihuan
  Chen}, \bibinfo{person}{Rongfan Li}, {and} \bibinfo{person}{Xin Peng}.}
  \bibinfo{year}{2021}\natexlab{a}.
\newblock \showarticletitle{A hybrid code representation learning approach for
  predicting method names}.
\newblock \bibinfo{journal}{\emph{J. Syst. Softw.}}  \bibinfo{volume}{180}
  (\bibinfo{year}{2021}), \bibinfo{pages}{111011}.
\newblock


\bibitem[\protect\citeauthoryear{Zhang, Hong, Zhang, Wan, Liu, and Sui}{Zhang
  et~al\mbox{.}}{2021b}]%
        {ZhangHZWLS21}
\bibfield{author}{\bibinfo{person}{Jingfeng Zhang}, \bibinfo{person}{Haiwen
  Hong}, \bibinfo{person}{Yin Zhang}, \bibinfo{person}{Yao Wan},
  \bibinfo{person}{Ye Liu}, {and} \bibinfo{person}{Yulei Sui}.}
  \bibinfo{year}{2021}\natexlab{b}.
\newblock \showarticletitle{Disentangled Code Representation Learning for
  Multiple Programming Languages}. In \bibinfo{booktitle}{\emph{Proceedings of
  ACL/IJCNLP}}. \bibinfo{pages}{4454--4466}.
\newblock


\bibitem[\protect\citeauthoryear{Zhou, Liu, Siow, Du, and Liu}{Zhou
  et~al\mbox{.}}{2019}]%
        {ZhouLSD019}
\bibfield{author}{\bibinfo{person}{Yaqin Zhou}, \bibinfo{person}{Shangqing
  Liu}, \bibinfo{person}{Jing~Kai Siow}, \bibinfo{person}{Xiaoning Du}, {and}
  \bibinfo{person}{Yang Liu}.} \bibinfo{year}{2019}\natexlab{}.
\newblock \showarticletitle{Devign: Effective Vulnerability Identification by
  Learning Comprehensive Program Semantics via Graph Neural Networks}. In
  \bibinfo{booktitle}{\emph{Proceedings of NeurIPS}}.
  \bibinfo{pages}{10197--10207}.
\newblock


\end{thebibliography}

%%
%% If your work has an appendix, this is the place to put it.
% \appendix

% \section{Research Methods}

% \subsection{Part One}

% Lorem ipsum dolor sit amet, consectetur adipiscing elit. Morbi
% malesuada, quam in pulvinar varius, metus nunc fermentum urna, id
% sollicitudin purus odio sit amet enim. Aliquam ullamcorper eu ipsum
% vel mollis. Curabitur quis dictum nisl. Phasellus vel semper risus, et
% lacinia dolor. Integer ultricies commodo sem nec semper.

% \subsection{Part Two}

% Etiam commodo feugiat nisl pulvinar pellentesque. Etiam auctor sodales
% ligula, non varius nibh pulvinar semper. Suspendisse nec lectus non
% ipsum convallis congue hendrerit vitae sapien. Donec at laoreet
% eros. Vivamus non purus placerat, scelerisque diam eu, cursus
% ante. Etiam aliquam tortor auctor efficitur mattis.

% \section{Online Resources}

% Nam id fermentum dui. Suspendisse sagittis tortor a nulla mollis, in
% pulvinar ex pretium. Sed interdum orci quis metus euismod, et sagittis
% enim maximus. Vestibulum gravida massa ut felis suscipit
% congue. Quisque mattis elit a risus ultrices commodo venenatis eget
% dui. Etiam sagittis eleifend elementum.

% Nam interdum magna at lectus dignissim, ac dignissim lorem
% rhoncus. Maecenas eu arcu ac neque placerat aliquam. Nunc pulvinar
% massa et mattis lacinia.

\end{document}